\documentclass[10pt,journal,compsoc]{IEEEtran}
\usepackage{cite}
\usepackage{xcolor}
\usepackage{ulem}

\ifCLASSINFOpdf
  \usepackage[pdftex]{graphicx}
\else
\fi
\begin{document}
\title{Leveraging Tendon Vibration to Enhance Pseudo-Haptic Perceptions in VR}

\author{Yutaro~Hirao,~%~\IEEEmembership{Member,~IEEE,}
        Tomohiro~Amemiya,~%
        Takuji~Narumi,~%~\IEEEmembership{Fellow,~OSA,}
        Ferran~Argelaguet,~%~\IEEEmembership{Fellow,~OSA,}
        and~Anatole~Lécuyer %~\IEEEmembership{Life~Fellow,~IEEE}% <-this % stops a space
\IEEEcompsocitemizethanks{\IEEEcompsocthanksitem Y. Hirao, T. Amemiya and T. Narumi are with the University of Tokyo, Tokyo 113-8654, Japan. E-mail: \{hirao, amemiya, narumi\}@cyber.t.u-tokyo.ac.jp.
\IEEEcompsocthanksitem A. Lécuyer and F. Argelaguet work with Univ. Rennes, Inria, IRISA, CNRS, Rennes, France.}% <-this % stops an unwanted space
\thanks{}}

%if necessarily
%E-mails: \{ferran.argelaguet, anatole.lecuyer\}@inria.fr.


%CHECK
\markboth{Journal of}% %Class Files,~Vol.~14, No.~8, August~2015}%
{Hirao \MakeLowercase{\textit{et al.}}: Vibrator Feedback to Enhance Pseudo-Haptics}



\IEEEtitleabstractindextext{%
\begin{abstract}
Pseudo-haptic techniques are used to modify haptic perception by appropriately changing visual feedback to body movements. Based on the knowledge that tendon vibration can affect our somatosensory perception, this paper proposes a method for leveraging tendon vibration to enhance pseudo-haptics during free arm motion. Three experiments were performed to examine the impact of tendon vibration on the range and resolution of pseudo-haptics. The first experiment investigated the effect of tendon vibration on the detection threshold of the discrepancy between visual and physical motion. The results indicated that vibrations applied to the inner tendons of the wrist and elbow increased the threshold, suggesting that tendon vibration can augment the applicable visual motion gain by approximately 13\% without users detecting the visual/physical discrepancy. Furthermore, the results demonstrate that tendon vibration acts as noise on haptic motion cues. The second experiment assessed the impact of tendon vibration on the resolution of pseudo-haptics by determining the just noticeable difference in pseudo-weight perception. The results suggested that the tendon vibration does not largely compromise the resolution of pseudo-haptics. The third experiment evaluated the equivalence between the weight perception triggered by tendon vibration and that by visual motion gain, that is, the point of subjective equality. The results revealed that vibration amplifies the weight perception and its effect was equivalent to that obtained using a gain of 0.64 without vibration, implying that the tendon vibration also functions as an additional haptic cue. Our results provide design guidelines and future work for enhancing pseudo-haptics with tendon vibration.

\end{abstract}

}

\maketitle

\IEEEdisplaynontitleabstractindextext

% \fi
\IEEEpeerreviewmaketitle

\IEEEraisesectionheading{\section{Introduction}\label{sec:introduction}}


\IEEEPARstart{I}{n} recent years, virtual reality (VR) technology has made remarkable progress, enabling users to feel as if a virtual object is actually present by providing visual and audio information. However, haptic information still has a narrower range of expression than audio-visual information, and various approaches have been proposed to present haptic information. Of these approaches, the pseudo-haptic technique is an interesting alternative for generating haptics; it is a method of modifying haptic perception by appropriately changing the visual feedback to body movement \cite{Pusch2011, Ujitoko2021}. Typical examples of pseudo-haptics techniques involve modifying the force of a spring \cite{Lecuyer2000} or perceived weight of an object \cite{Taima2014, rietzler2018breaking, Samad2019} by changing the control-display gain to the actual motion in a virtual environment. The main advantage of the pseudo-haptics technique is that it can present haptic perceptions primarily through visual stimuli and does not require bulky equipment. However, the perceived force intensity presented by conventional techniques of pseudo-haptics is severely limited because the pseudo-haptic technique separates visual and haptic information, and discrepancies that are too large induce discomfort or break the haptic perception \cite{Pusch2011, rietzler2018breaking, honda2013imposed}.

Some studies have attempted to improve the range of acceptable motion alteration by increasing the contribution of visual information to sensory integration by expanding the field of view of a head-mounted display (HMD) \cite{williams2019estimation} or using a more human-like avatar \cite{ogawa2020effect}. However, modifying the haptic perception by directly increasing the contribution of visual information is limited. In this study, considering that tendon vibration affects somatosensory perception, we investigate the potential of tendon vibration for expanding the range of pseudo-haptic perceptions.

\begin{figure}[t]
 \centering
 \includegraphics[width=\columnwidth]{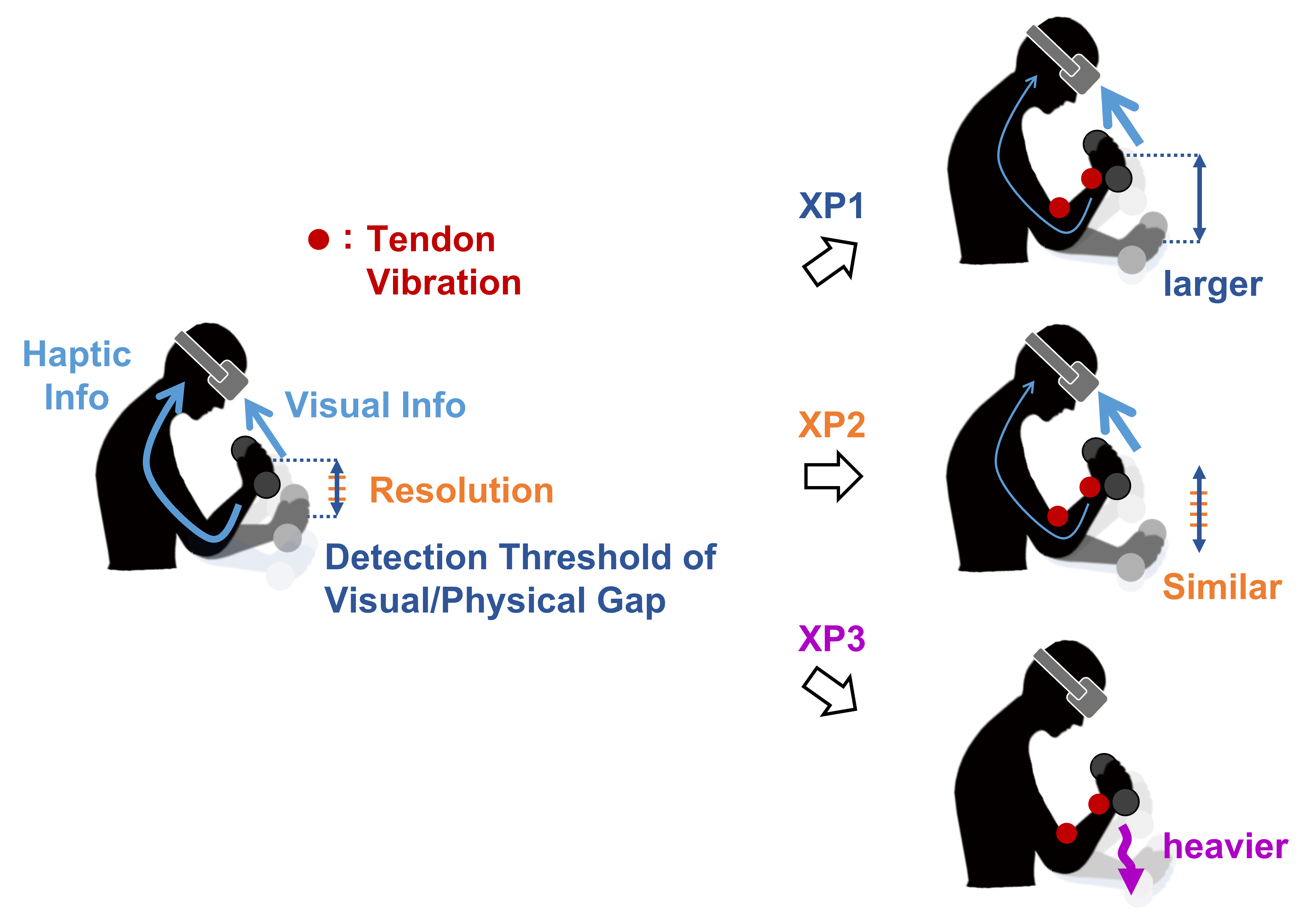}
 \caption{The figure shows the summary of results: Tendon vibration was suggested to function as noise on somatosensory information, making users rely more on visual information and leading to a potentially wider range of applicable visual motion gain (XP1). Moreover, the tendon vibration does not change the resolution of the pseudo-weight perception (XP2). Furthermore, tendon vibration has a side effect, increasing weight perception (XP3).}
 \label{SummaryOfPaper}
\end{figure}

Vibration has several possible effects on somatosensation. When a vibration is applied to a muscle or tendon, the primary afferent of the muscle spindle is activated. This can produce illusions of position, motion, and force \cite{proske2012proprioceptive, taylor2017muscle, proske2019neural}. In addition to these illusions, the tonic vibration reflex (TVR) occurs, which causes sustained contraction of the vibrated muscle and inhibition of the activity of the antagonist muscle \cite{hagbarth1966motor}. Although many studies confirmed the effect of tendon vibration on somatosensory information, it is also said that the effect is not robust and the direction of the motion/force illusion can be different for each user, situation, or user motion. Then, based on these findings, we hypothesized that the tendon vibration can be used as noise on somatosensory information during user motion and can make users rely more on visual information, thus enhancing the effect of pseudo-haptics.
In this study, we evaluate the effect of tendon vibration on pseudo-haptic weight perception during free arm motion and discuss design guides for combining tendon vibration and pseudo-haptics techniques. As a systematic investigation of the effect, we conducted three experiments investigating the potential effect of tendon vibration on the range and resolution of pseudo-weight perception. Figure \ref{SummaryOfPaper} provides a brief summary of results. The first experiment investigated the effects of tendon vibration on the agonist muscle, antagonist muscle, or both on the detection threshold (DT) of the discrepancy between physical and virtual motion with a motion gain. Because users may begin to feel discomfort when they recognize the discrepancy \cite{Pusch2011}, this DT can be considered the strictest maximum applicable motion gain for generating pseudo-weight perception. Then, the second experiment investigated the effect of tendon vibration to the agonist muscle on the just noticeable difference (JND) in pseudo-weight perception induced by motion gain. This JND indicates the resolution of pseudo-weight perception. We consider that the DT and JND can be used to discuss whether tendon vibration improves the range of presentable pseudo-weight perception. Finally, the third experiment investigated the side effect of tendon vibration on weight perception. This concerns the point of subjective equality (PSE) of weight perception induced by tendon vibration, compared with pseudo-weight perception induced by motion gain. These three experiments cover the effects of combining tendon vibration and motion gain on pseudo-weight perception in VR.

The remainder of this paper is structured as follows. Section 2 introduces related work. We introduce the theoretical background of pseudo-haptics and effects of tendon vibration on somatosensory  perception. Sections 3 to 5 describe the three experiments. Then, Section 6 provides a general discussion of all experimental results and comprehensive discussion of the effect of tendon vibration on the pseudo-haptic technique and design guide for application. Finally, Section 7 concludes the paper.

\section{Related Work}
\subsection{Theoretical Background of Pseudo-Haptics}

\begin{figure}[b]
 \centering
 \includegraphics[width=\columnwidth]{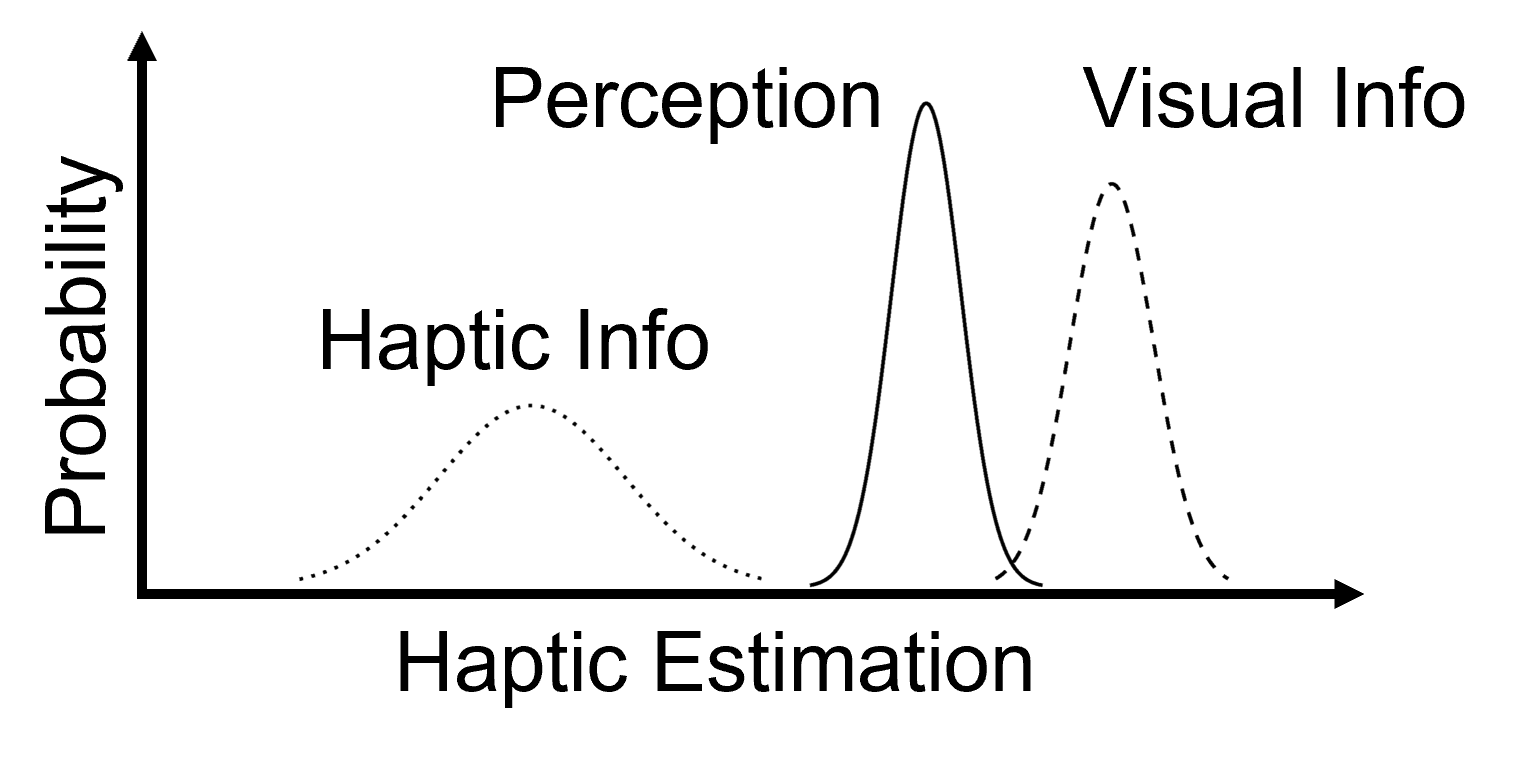}
 \caption{Concept figure of the maximum likelyhood estimation of the sensory integration process. It shows that the haptic estimation (final perception) is made by the integration of visual and haptic information considering their reliability, i.e., the variance of the information. Note that the distribution and mean of the visual and haptic info are only for illustrative purposes.}
 \label{MLE}
\end{figure}

The pseudo-haptic technique is used to induce haptic illusions by manipulating visual feedback to user movement \cite{Pusch2011, Ujitoko2021}. It can be applied for many properties of haptic perceptions, such as compliance \cite{Lecuyer2000, Argelaguet2013, takamuku2015you}, weight \cite{honda2013imposed, Taima2014, rietzler2018breaking, Samad2019}, and friction \cite{Narumi2017, Ujitoko2019, ujitoko2019modulating, lecuyer2004feeling, hachisu2011virtual}. The mechanism of pseudo-haptics has not yet been fully revealed. Still, several theories have been proposed and discussed. One of the most conventional explanation is that visual and haptic information are integrated based on the maximum likelihood estimation (MLE) \cite{ernst2002humans, ernst2004merging}. This theory explains that sensory information is integrated based on its reliability (likelihood) for people to estimate the properties of physical objects (Fig. \ref{MLE}). Ernst and Banks confirmed this in the estimation of the width of an object when the information of width is different between vision and haptics \cite{ernst2002humans}. In particular, MLE well describes spatial perception. In addition, considering the relationship between motion and force, another explanation of the mechanism of pseudo-haptics is based on the indication that the human central nervous system performs forward dynamics calculations (FDC) \cite{wolpert1995internal, ariff2002real, honda2013imposed} and inverse dynamics calculations (IDC) \cite{shadmehr1994adaptive, runeson1983kinematic, takamuku2015you}. In FDC, the motion of a target object is estimated from the force applied to it. Honda et al. adjusted the visual delay and mass of a manipulandum in a system where the cursor on a monitor can be manipulated by the manipulandum. They confirmed that the sensory-motor prediction error owing to the visual delay was misattributed to the mass estimation \cite{honda2013imposed}. Moreover, in IDC, the force applied to a target object is inferred from its motion. Takamuku and Gomi investigated the intensity of motion resistance while varying the visual motion of a cursor on a screen to the periodic motion of a stylus pen \cite{takamuku2015you}. They confirmed that the intensity of motion resistance associated with cursor delay correlates with the acceleration of the cursor in the direction of movement. This suggests that their subjects used IDC with visual motion information as input to form an internal model of the dynamics for the interaction (in this case, the spring-damper system). By further developing these theories that can well explain the primary bottom-up process of the sensory integration or relationship between motion and force estimation, the Bayesian theory was proposed; it considers top-down influences such as the intervention of predictions/prior-knowledge, and complement/replacement of sensory input by individual memory/experience \cite{Pusch2011, ernst2004merging, knill2004bayesian}. Moreover, a more unified explanation is obtained using the free energy principle, which considers the loop between perception and behavior change \cite{friston2009free}.

\subsection{Limits of Pseudo-Haptics}
As discussed, the manipulation of various haptic perceptions is possible by only adjusting visual feedback. However, the pseudo-haptic technique is not omnipotent. Pseudo-haptics separate visual and haptic information. Here, larger gaps are necessary to induce more intensive perception; however, gaps that are too large result in discomfort or a failure to integrate the information \cite{Pusch2011, rietzler2018breaking, honda2013imposed}. Based on Pusch and L{\'e}cuyer's model \cite{Pusch2011}, larger discrepancies could result in larger complements and substitutions of sensory information from personal memories and experiences,  which would increase individual differences. Moreover, if the gap is too large, the information is determined to be coming from a different source and is not integrated \cite{ernst2004merging}. While, FDC and IDC theory can provide another explanation in that the illusion would not arise when the brain no longer has a model to which the large gap can be attributed \cite{honda2013imposed}. 

Few studies have been conducted to solve the challenges of pseudo-haptic techniques. Ban and Ujitoko proposed a method for inducing pseudo-haptics more effectively by connecting the discrepancy between a user's finger and the cursor on a touchscreen with a visual string \cite{ban2018enhancing}. Another possible approach involves increasing the reliability of visual information in sensory integration, i.e., to decrease the variance of visual information in MLE so that the final perception is more dependent on visual information. Williams and Peck confirmed that wider fields of view in VR could extend the applicable visual gain for redirected walking techniques \cite{williams2019estimation}. Furthermore, Ogawa et al. concluded that more realistic avatars could make participants notice the discrepancy between virtual and physical hand movements less than the abstract avatar (a sphere object) \cite{ogawa2020effect}. These studies did not reference pseudo-haptic feedback; however, their targeting technique, i.e., retargeting technique, leads to visual and physical discrepancies. As such, their findings can be extended to pseudo-haptic feedback. Nonetheless, these approaches do not manipulate the haptic information, thus limiting their efficacy. Accordingly, we investigated a method that alters haptic feedback using simple equipment, specifically through tendon vibration stimulation, to address this issue.

\subsection{Effect of tendon vibration on somatosensory information and pseudo-haptic perception}
\label{LR:TV}
\begin{figure}[b]
 \centering
 \includegraphics[width=\columnwidth]{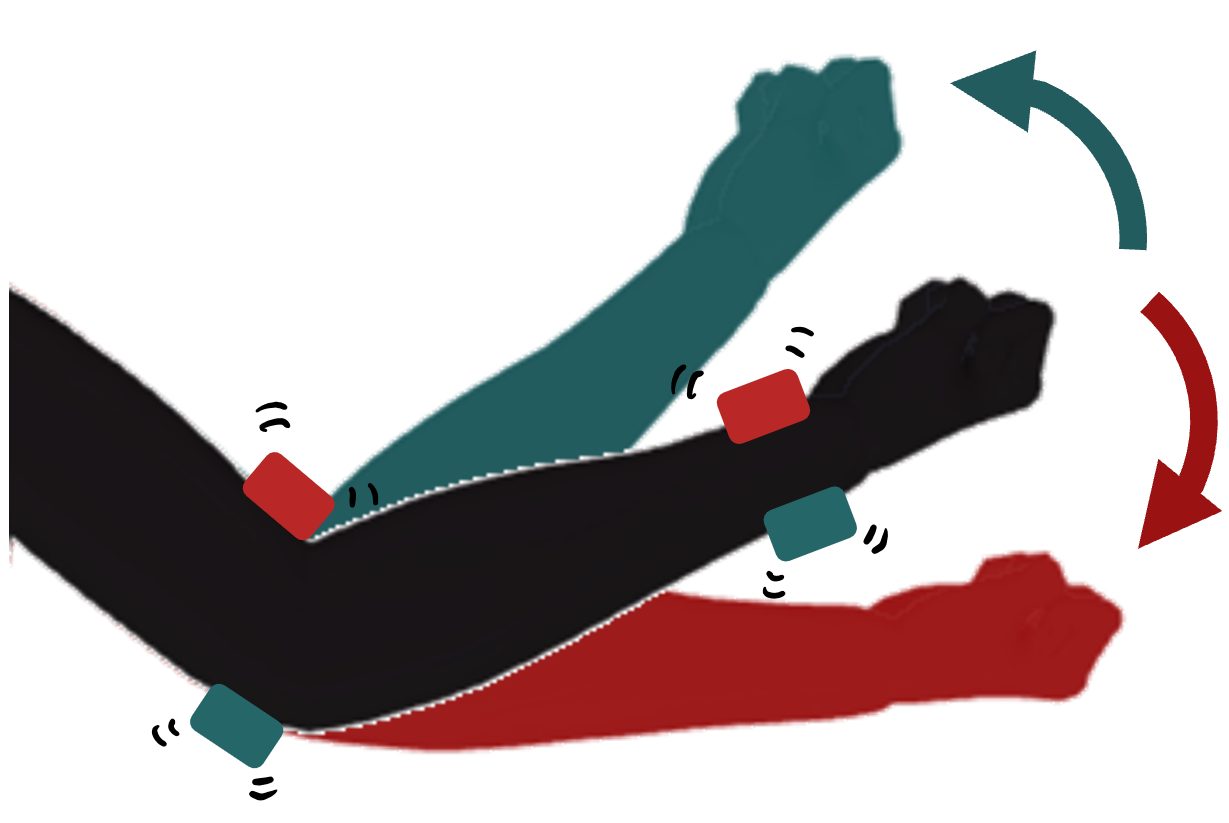}
 \caption{Relationship between the position of tendon vibration feedback and direction of motion illusion.}
 \label{TendonVibrationIllusion}
\end{figure}

When vibration is applied to a tendon or muscle, it activates the primary afferents of the muscle spindle. The muscle spindle is a sensor of position and motion \cite{matthews1972mammalian}, so the vibration can create the illusion of movement in the direction of the muscle stretch (Fig. \ref{TendonVibrationIllusion}) \cite{goodwin1972contribution}. This vibration also induces a tonic vibration reflex (TVR) that contracts muscles \cite{hagbarth1966motor}. When we resist the motion and try to remain static, a motion illusion is produced in the opposite direction of the TVR. Tendon vibration can also affect force and weight perception (e.g. \cite{mccloskey1974estimation, jones1985effect}). The optimal vibration parameters for generating the somatosensory illusion vary within studies, but a vibration frequency of 70 to 100 Hz and an amplitude of approximately 5 g is deemed sufficient and widely used for VR experiments (e.g., \cite{taylor2017muscle, le2020influence, ushiyama2021modulation}). The required minimum duration of vibration to induce the illusion also varies, but several studies have indicated that a minimum of 6 seconds is required \cite{naito1999illusory, kitada2002perceptual}.

Although many tendon vibration studies have been conducted for static situations, several studies have confirmed that the tendon vibration illusion can also be induced during active motion \cite{sittig1987contribution, inglis1990effect}. For instance, during elbow flexion-extension movements, applying vibration to the biceps affects position perception in slow movements and velocity perception in fast movements \cite{sittig1987contribution}. Moreover, vibration on the biceps (agonist muscle) does not affect positional accuracy during arm flexion-extension movements of 40-60 deg/s, whereas vibration on the triceps (antagonist muscle) does, suggesting that muscle spindle afferents from the extending antagonist muscle contribute to limb positional accuracy during voluntary movements \cite{inglis1990effect}. However, the reproducibility of the effect is inconsistent across studies, mainly because vibrating one muscle may induce some activity in neighboring or antagonistic muscles, resulting in a change in the direction of the illusion or even its disappearance \cite{burke1976responses, le2020influence}. Furthermore, individual differences in anatomy increase the difficulty in selectively vibrating the intended muscle. Moreover, the tendon vibration illusion is strongly influenced by visual information (e.g. \cite{le2020influence, tsuge2012interaction, fusco2021visual}) and cognitive factors such as learning, attention, and imagination \cite{taylor2017muscle}. For example, neuroimaging studies have shown that the same brain regions are activated in both motor imagery \cite{porro1996primary} and illusory arm movements \cite{naito1999illusory}. Then, Thyrion and Roll confirmed that motor imagery weaken/strengthen the illusory motion induced by the vibration of biceps, and even modulate its perceived direction \cite{thyrion2009perceptual}. In this manner, the motion illusion of tendon vibration may not be always robust.

In addition to its impact on position and motion estimation, several investigations have explored the influence of vibration on force and weight estimation. The majority of studies have indicated that vibration of the contracting muscle enhances the sense of force or weight (e.g. \cite{cafarelli1981effect, jones1985effect, luu2011fusimotor}). The mechanism underlying the sense of force and weight is not yet fully comprehended, although somatosensory information is recognized to be a composite mechanism of central and peripheral signals. These signals include the afferent signal generated by external stimuli and the efferent copy created in corollary discharge with the command to execute physical movements. The efferent copy is occasionally described as the sense of effort, and both the sense of effort and peripheral signals are recognized to affect force and weight perception \cite{proske2012proprioceptive, proske2019neural}.

The reason for overestimating the exerted force through tendon vibration can be explained in terms of two aspects: afferent and efferent signals. Regarding afferent signals, vibrations to the tendon and TVR cause neurons in the Golgi tendon organ, which senses the degree of muscle contraction, to activate. These additional afferent signals increase force and weight perception \cite{cafarelli1981effect}. This is also supported by the reduction of the vibration effect during muscle fatigue \cite{cafarelli1986effect} and the decrease in weight perception due to the desensitization of the Golgi tendon organ caused by relatively long-term vibration \cite{burke1976responses, fallon2007vibration, luu2011fusimotor}.

Concerning efferent signals, vibration of the driving muscle leads to the inhibition of the antagonist muscle, increasing the effort required to overcome this inhibition and exert the target force. This results in increased force or weight perception \cite{jones1985effect, proske2019neural}. However, these effects are not always consistent, as some studies report conflicting results. McCloskey et al. observed that muscle exertion perceived in vibrated muscles is underestimated \cite{mccloskey1974estimation}. They suggested that participants do not consider the force from TVR as additional force input; thus, TVR resulted in reducing the sense of effort. Moreover, the effects vary depending on the context, such as with tasks or questionnaires \cite{mccloskey1973differences, monjo2018sensory}. Furthermore, as motion and force perception are closely related, the difficulty involved in using tendon vibration to control motion illusion implies that controlling force perception may also be challenging.

Given these findings, this paper endeavors to explore whether the impact of tendon vibration can be accurately modulated or if it can be employed as a source of noise in somatosensory information to enhance pseudo-haptics. Additionally, this paper aims to assess the impact of tendon vibration and pseudo-haptics on haptic perception by using the same scale, enabling the comparison of their effects and the design of haptic perception through their integration.

%%%%%%%%%%%%%%%%%%%%%%%%%%%%%%%%%%%%%%%%%%%%%%%%%%%%%%%%%%%%%%%%%%%%%%%%%%%%%%%%%%%%%%%%%%
%XP1
%%%%%%%%%%%%%%%%%%%%%%%%%%%%%%%%%%%%%%%%%%%%%%%%%%%%%%%%%%%%%%%%%%%%%%%%%%%%%%%%%%%%%%%%%%

%

\section{XP1: Influence of tendon vibration on the range of the maximum applicable visual/physical gap}
This experiment investigates the influence of tendon vibration on the applicable range of pseudo-weight perception. In this experiment, the DT of the discrepancy between actual and VR motion with a motion gain was investigated with the inner, outer, and both sides of vibration on the wrist and elbow tendons, as well as a with no vibration condition. As reviewed in Section \ref{LR:TV}, the effects of tendon vibration on somatosensory perception are complex. In particular, the experiments in previous studies were conducted in a strict experimental environment with the subject's body part fixed, as the purpose was to purely confirm the effects of tendon vibration on somatosensory perception. In contrast, this experiment aims to clarify the effects in a more practical situation where the arm is free to move,  considering consumer applications. The research question guiding this experiment was whether and how tendon vibration affects the DT of the visual/physical motion discrepancy in our setup.

\subsection{Methods}
\subsubsection{Apparatus}
\begin{figure}[b]
 \centering
 \includegraphics[width=\columnwidth]{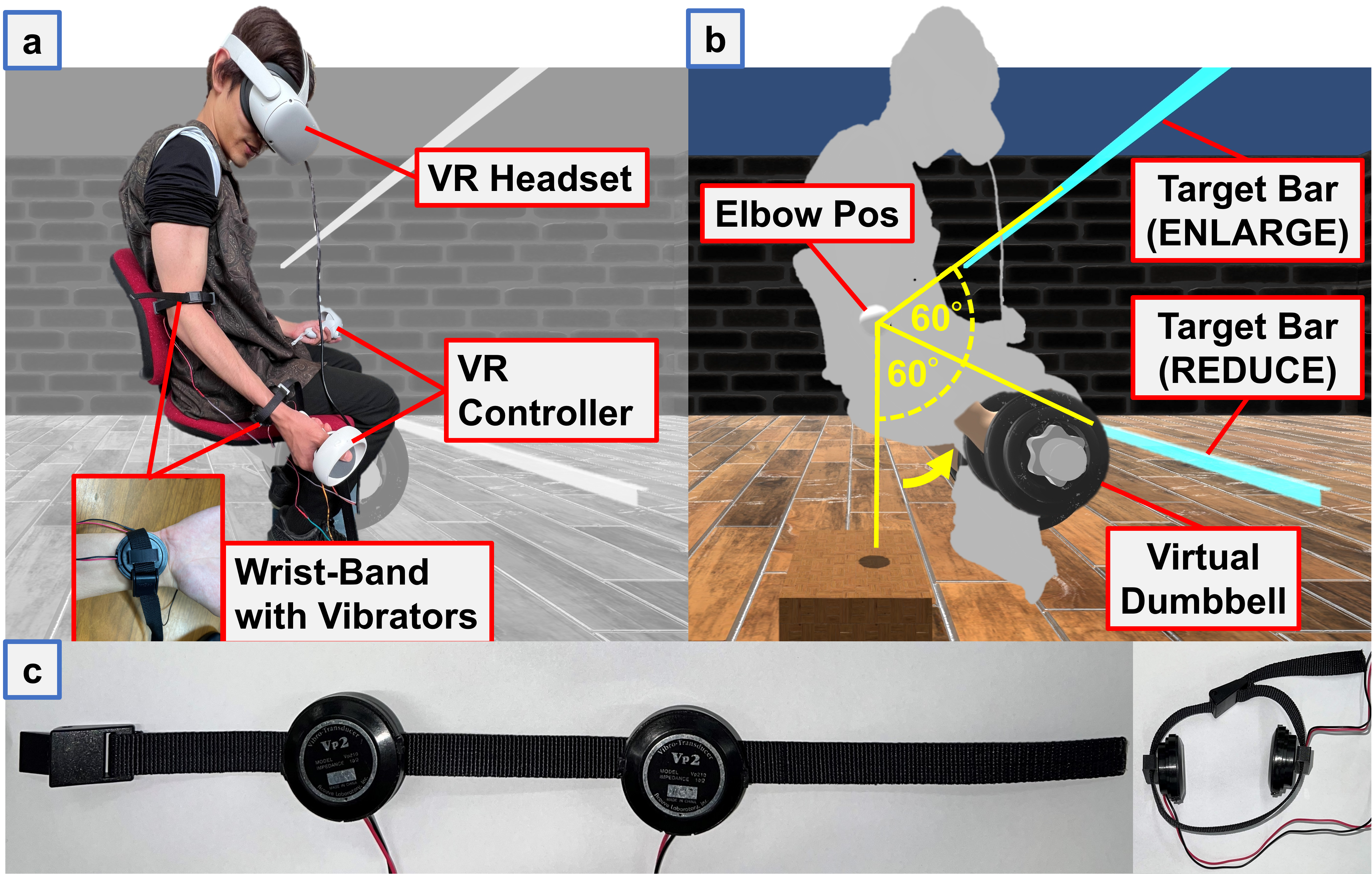}
 \caption{Apparatus (a), virtual environment (b), and wrist band with vibrators (c). Participants wore noise-cancelling headphones during the experiment. In addition, the two target bars were not presented simultaneously but one-by-one in each condition.}
 \label{Environment}
\end{figure}

Figure \ref{Environment}a  shows the experimental apparatus. The experiment was conducted with participants in a sitting position. Participants wore wristbands with two vibrators (VIBRO transducer VP210, Acouve Lab. Inc.), on the wrist and elbow of the right arm. The vibrators were positioned to stimulate the inner and outer tendons of the wrist and elbow, respectively. We would like to note that the reason of having vibrators not only on an elbow but on a wrist is that the position/force estimation is constructed by combining signals from multiple synergist muscles/tendons \cite{proske2012proprioceptive} and stimulating them can enhance the kinesthetic illusion\cite{ushiyama2020effects}. When the bands were wrapped around the subject's arm, it was wrapped with enough tightness so that the vibrators, which was in contact with the skin, did not move from its initial position during the experiment trial.
The participants wore a VR headset (Oculus Quest 2) and noise canceling headphones and held a VR controller in each hand. The frequency of the vibration was 80 Hz and the amplitude was about 5 g except when vibrating both the inner and outer tendons, for which it was about 2.5 g. These amplitudes was validated with an accelerometer for each vibrator.

\subsubsection{Conditions}
\label{xp1 condition}
The experiment used a within-subjects design. The tested condition related to the placement of the vibrations and included three levels: the inner, outer, and all tendons vibrating, including the control condition without vibration. The amplitude of the vibration was 5 g for the inner and outer tendons, and 2.5 g for the all tendons condition. Motion illusion is known to weaken when the amplitude of tendon vibration is reduced \cite{schofield2015characterizing}. However, we considered that it was more important to maintain the total intensity of vibration stimuli the same. This is because if the vibration simply induces a motion illusion, the effect of inner and outer tendon vibration should cancel each other.

\subsubsection{Procedure}
All the experiments in the paper were monitored by the local ethical committee in the University of Tokyo under protocol No. UT-IST-RE-230517-1.
Twenty participants (twelve males and eight females in their twenties) participated in the experiment. First, the objectives, methods, and procedures were explained to them. Next, the participants completed a consent form. Then, the participants were asked to wear all the equipment (Fig.\ref{Environment}a with headphones).

The experiment comprised two distinct parts. The first part aimed to calibrate the vibrators' position and evaluate the motion illusion induced by tendon vibration; and the second part is the main part of this experiment, where the DT is measured using the staircase method.

\textbf{\textit{Calibration and measurement of motion illusion induced by tendon vibration:}} Participants were initially informed that the vibration could induce a motion illusion, without specifying its direction to avoid bias. Then, participants sat with their right arm extended straight down and with their eyes closed as they looked straight ahead. Next, an experimenter activated the vibration stimuli to the inner sides of the participant's right wrist and elbow for six seconds. Subsequently, the experimenter inquired whether the participants felt a motion illusion during the vibration. Previous studies have demonstrated that six seconds is the minimum duration required to induce motion illusion \cite{naito1999illusory, kitada2002perceptual}, which was also confirmed by our informal pilot test. If the participants did not experience the motion illusion, the experimenter slightly adjusted the vibrator's position, repeating the process five times or until the motion illusion was perceived. In cases where participants did not experience the motion illusion after the second and fourth repetitions, the experimenter activated a series of vibrations, following the order of the inner, outer, and inner tendon, to enhance the probability of participants experiencing the illusion \cite{kito2006sensory}. The same calibration process was repeated for the vibration of the outer tendon.

After calibrating all vibrator positions, the participants were asked to quantify the vector of motion illusion they experienced. Initially, the experimenter provided inner vibration stimuli to the participants for six seconds. The participants were then asked to replicate the motion they experienced using the right controller. In particular, participants pulled the index trigger at the initial position, replicating the motion illusion with holding the trigger until the end of the motion illusion. A "beep" sound was provided when the trigger was pulled or released. This replication task was performed three times for each inner and outer tendon vibration. It is important to note that participants remained with their eyes closed throughout the task. After the first part, the position of the elbow was measured to co-localize the user with the virtual content. The participants were instructed not to move their elbow through the second part.

\textbf{\textit{Investigation of DT of visual/physical discrepancy:}} In short, the second part involved performing a lifting task with a virtual dumbbell wherein a visual gain was applied to the motion and answering if the participants noticed discrepancies between the actual and visual motions. The staircase method was used in the second part. Before the main trials, participants conducted a training phase to understand the staircase procedure without vibrations or visual gain. 

The visual gain was categorized into two groups, ENLARGE and REDUCE, with ENLARGE having a visual gain greater than 1.0, indicating that the virtual motion was enlarged compared to the actual motion, and REDUCE having a visual gain less than 1.0, indicating that the virtual motion was reduced compared to the actual motion. Therefore, there were eight blocks consisting of four vibration conditions with the ENLARGE and REDUCE visual gain groups. Each block had two series of visual gain and the initial motion gain for ENLARGE was 1.0 or 2.0, while for REDUCE, it was 0.5 or 1.0. The gain was adjusted by one step from ten levels within each group, depending on the participants' responses to a questionnaire asking whether they experienced the motions of virtual and physical hands were the same or different. When participants answered "the same," the visual gain moved one step farther from 1.0, and when they answered "different," it moved one step closer to 1.0. Each series of the staircase method was presented in turn and ended after five turning points of changing the direction of the visual gain. Furthermore, the visual gain did not cross over the initial visual gain, that is, did not go under the lower initial gain and not go over the higher initial gain. If the answer would make the visual gain cross over the initial gain, the visual gain remained at the value, and the count of turning points was increased. The order of the vibration conditions was counterbalanced between participants, and the other orders on ENLARGE/REDUCE and the two series of visual gain were randomly chosen.  

A detailed explanation of the lifting task is as follows. Figure \ref{Environment}b shows the virtual environment. The VR environment depicted a virtual dumbbell placed at the participant's right hand where their right arm would naturally reach when extended straight down in a seated position. Participants could see a virtual right hand at the position of their right hand through the VR headset, and they could grab or release it by pulling or releasing the index trigger of the VR controller. Throughout the task, white noise was presented to the participants to mitigate the auditory interference caused by the vibrators. First, participants grabbed a dumbbell and waited for 6 seconds before lifting it. Here, if the current condition was the condition with vibration, the vibration presentation started. The 6 seconds pre-vibration was intended to more certainly induce the illusion \cite{naito1999illusory, kitada2002perceptual}. Moreover, a 6 seconds countdown display and metronome sound at 120 bpm were simultaneously presented. After the 6 s, the participants lifted the virtual dumbbell until it touched the blue bar in front of them in 2 seconds. Here, the participants were instructed to use the metronome sound for measuring the 2 seconds. Moreover, the "2 seconds" was determined to be a sufficient duration to have some effect and let the participants move with moderate speed by our pilot test, where we compared 1, 2, and 3 seconds in the same setup as this experiment. Participants were instructed to bend their elbow when they lifted the dumbbell. 

A gain was applied to the rotation of the elbow. The blue bar was placed at 60 degrees for REDUCE and 120 degrees for ENLARGE from the initial posture of the right arm. The position of the blue bar was determined so that the participants were required to physically perform the same range of arm motion to reach the blue bar for the ENLARGE and REDUCE conditions in the stair-case method. Specifically, in ENLARGE condition, the participants need to physically move their arm 60 degrees when the gain is 2.0 and 120 degrees when the gain is 1.0 while in REDUCE condition, they need to move their arm 60 degrees when the gain is 1.0 and 120 degrees when the gain is 0.5. Note that only one blue bar was presented during the task, not simultaneously as shown in the Fig. \ref{Environment}b. When the virtual dumbbell reached the blue bar, all virtual objects and stimuli disappeared. Then, a virtual panel with the written question "I felt that the motions of my virtual and physical hand were:" "the same" or "different," appeared in front of the participants. The participants were asked to choose one of the two options using the left controller.

During the trials, participants could take a break whenever and as often as they wanted. After finishing all trials, the participants removed all their equipment and completed the free-structured interview. The participants were remunerated with an Amazon gift card of 20 euro for their participation.

\subsubsection{Collected data}
The vector of the motion illusion induced by the vibrations were measured in the first part. In addition, the values of motion gain at the turning point of the staircase method were measured in the second part. These values were averaged to calculate the DT. In the end of the experiment, participants were asked to fill in a free structure interview.

\subsubsection{Research Question}
We had one research question regarding the impact of tendon vibration on the DT:

\begin{itemize}[\IEEEsetlabelwidth{\textrm{[Q1-1]}}]
    \item [\textrm{[Q1-1]}] Does tendon vibration effect the detection threshold, and if so, how does it function?
\end{itemize}

\begin{figure}[t]
 \centering
 \includegraphics[width=\columnwidth]{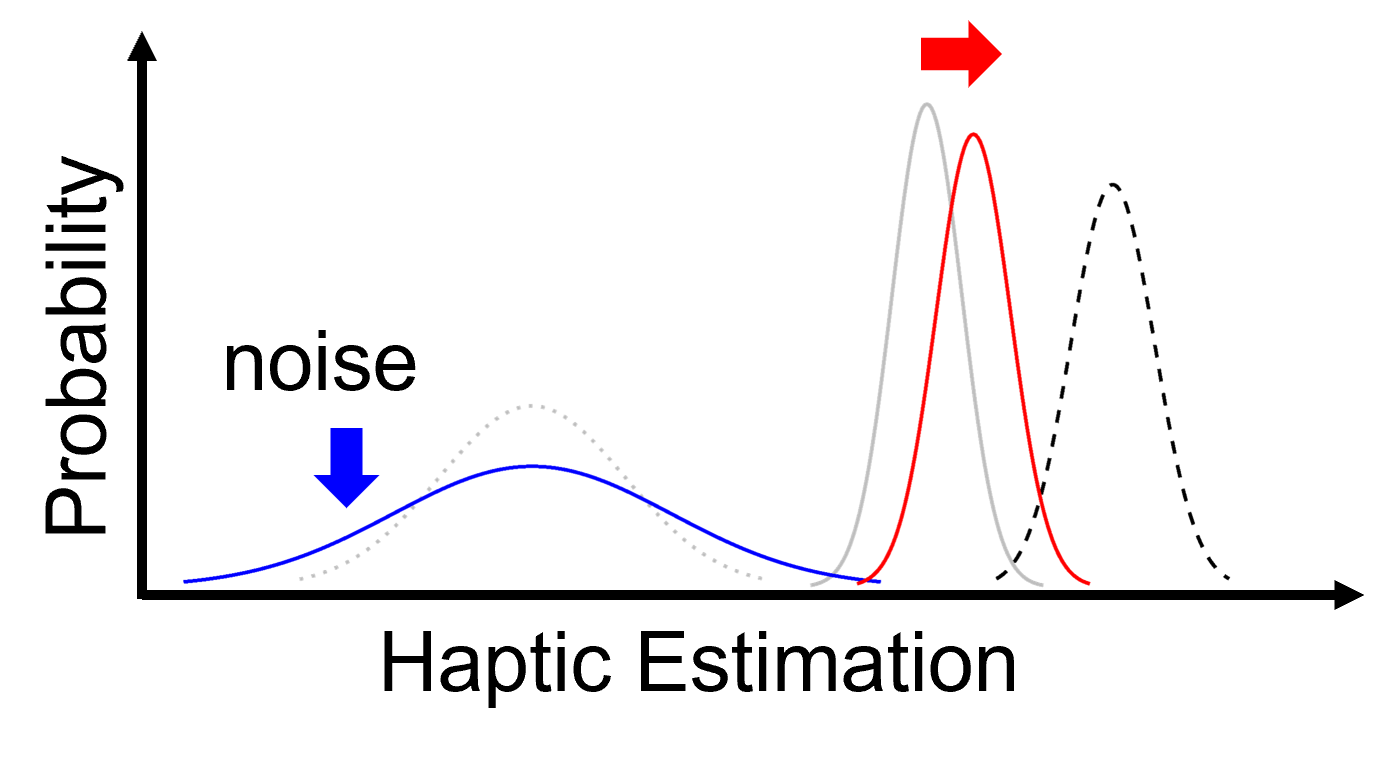}
 \caption{The figure shows the concept figure of MLE with noise added on haptic information. Noise on haptic information increases its variance and shifts the perception to the value estimated by visual information.}
 \label{MLE_noise}
\end{figure}

Considering the complexity of the effect of tendon vibration reviewed in the section \ref{LR:TV}, we think that there are two possibilities of how tendon vibration functions for the pseudo-haptics. The first possibility is that tendon vibration can purely induce the intended muscle spindle input, leading to the intended motion illusion. As the activation of muscle spindle induces the perception of the extension of muscle, this possibility implies that the DT of the visual/physical motion discrepancy becomes higher/lower with the inner/outer tendon vibration than that without vibration and both sides' vibration. Whereas, the second possibility is that tendon vibration functions as noise on somatosensory sensation. The tendon vibration was confirmed to degrade proprioceptive responsiveness \cite{bock2007method, brun2015anchoring}. Here, noise on sensory information can increase its variances and thus increases the contribution ratio of the other sensory information in the sensory integration process \cite{ernst2002humans}. In this case, if the tendon vibration functions as noise on somatosensory sensation, the visual information should be more dominant for position perception, and users will notice the discrepancy between visual and somatosensory information less (Fig. \ref{MLE_noise}). Therefore, this second possibility means that the visual gain of the DT becomes higher/lower with vibration conditions than that without vibration condition for the ENLARGE/REDUCE visual gain group.

\subsection{Results}

\subsubsection{Direction and amount of motion illusion at rest}
All participants successfully confirmed to experience motion illusion in the first part of the experiment. The data of one participant was removed because of recording issues and therefore the data of nineteen participants was used to analyze the results of the first part. Motion illusion was measured as a vector from the initial position to the end position of the illusion. Then, we first assigned 1 or -1 for all data of the motion illusion to observe the relationship between the inner/outer tendon vibration and direction of the motion illusion.  In particular, 1 was assigned for the results of motion illusion if the arm moved in a folding direction (forward), and -1 was assigned for the opposite. Here, the ratio of 1s to -1s were 13:6 for inner vibration and 16:3 for outer vibration. Then, the Spearman rank correlation test was conducted for the direction results of the motion illusion between inner and outer tendon vibration. Consequently, we could not determine a significant correlation. 

In addition, we analyzed the effect of tendon vibration on the amount of motion illusion. Here, we used the absolute value of the results of the motion illusion. The average of the absolute distance was $0.19 \pm 0.23$ (m) and $0.16 \pm 0.25$ (m) in the inner and outer tendon vibration, respectively. We conducted the Pearson correlation test for the results of the absolute distance of the motion illusion between inner and outer tendon vibration. Here, we determined a significant positive correlation between the absolute distances of motion illusion of the inner and outer tendon vibration ($r=0.74, p<0.001$).

\subsubsection{DT of the visual/physical motion discrepancy} 
\begin{figure}[t]
 \centering
 \includegraphics[width=\columnwidth]{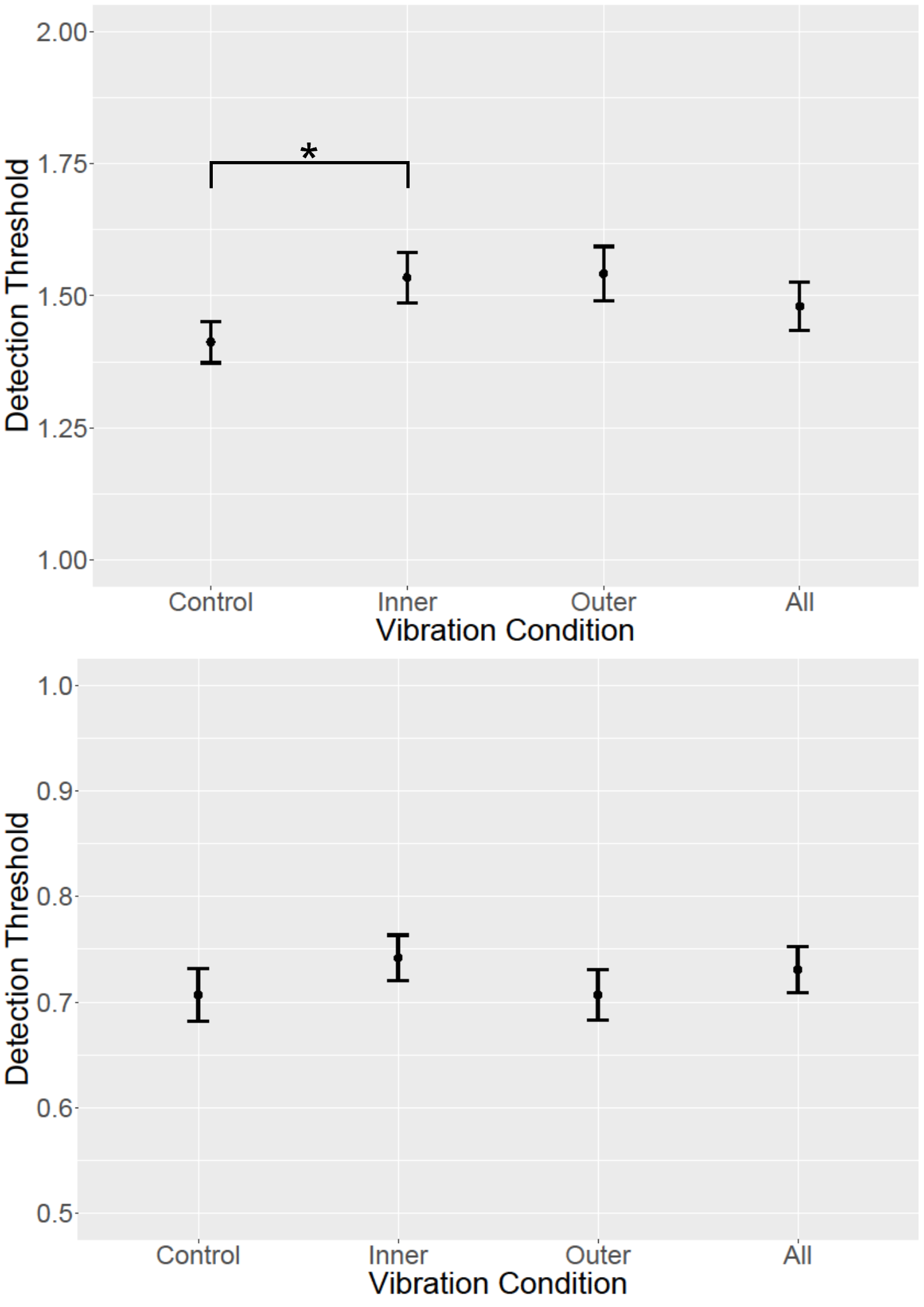}
 \caption{The figures show the average of the DT of the discrepancy between physical and virtual motion with a visual gain. The top figure shows the results of the ENLARGE group, and the bottom figure shows the REDUCE group. The error bars indicate the standard error. "*" indicates the statistical difference with $p < 0.05$.}
 %and "+" indicates a trend with $p < 0.1$.}
 \label{DT_Combined}
\end{figure}

Figure \ref{DT_Combined} shows the results of the DT. To obtain the DT within a block of the staircase method, ten turning points from both series of staircase methods were averaged. In the statistical analysis, we used the inverse values of the REDUCE group to enable a comparison of the effect of vibrations on the DT between the ENLARGE and REDUCE groups. Two-way analysis of variance (ANOVA) was conducted on the vibration conditions vs. ENLARGE/REDUCE group, and when the sphericity assumption was violated (Mauchly’s sphericity test), the degrees of freedom were adjusted using the Greenhouse-Geisser correction. A significant main effect was observed in the ENLARGE/REDUCE group factor ($F(1, 19)=4.56, p=0.0460, \eta^2=0.0324$), and a trend of interaction effect was also determined ($F(2.54, 48.17)=2.84, p=0.0562, \eta^2=0.0288$). Then, Shaffer’s modified sequentially rejective Bonferroni post-hoc tests were conducted for multiple comparisons. Consequently, a significant difference between control ($mean=1.41, \sigma=0.17$) and inner ($mean=1.53, \sigma=0.21$) vibration condition ($p=0.0498, r=0.56$)
%, and a trend between control and outer ($mean=1.54, \sigma=0.23$) vibration condition ($p=0.054, r=0.51$) 
in High visual gain were found. Moreover, as a result of other post-hoc test on the simple effect for the interaction, we determined a significant difference between the ENLARGE and inverse REDUCE group for the inner vibration condition ($p=0.0165, \eta^2=0.142$).
%and a trend at Outer condition ($p=0.0675, \eta^2=0.0461$).

\subsubsection{Relationship between motion illusion and DT}
First, we analyzed the relationship between the DT results and direction of motion illusion measured in the first part of the experiment. Because we focused on the direction and its effect on the DT, we assigned 1 or -1 for all data of the motion illusion and DT. In particular, 1 was assigned to the DT results if it increased compared with the DT without vibration, and -1 for the opposite results. Then, the Spearman rank correlation test was conducted for the pairs of the results of motion illusion and DT within the same tendon vibration condition. Consequently, we found a significant negative correlation between the results of motion illusion and the DT of the REDUCE group, with inner tendon vibration ($r=-0.46, p=0.047$).

In addition, we analyzed the relationship between the DT results and distance of motion illusion. To obtain the effect the vibration had on the DT, we standardized the DT data by the DT values without vibration. Subsequently, we performed the Pearson correlation test between the absolute distance of motion illusion and DT results pairs, and determined that there was no significant correlation.

\subsection{Discussion}

\subsubsection{Effect of tendon vibration on motion illusion at rest} 
Unlike previous studies (e.g. \cite{inglis1990effect, sittig1987contribution}), the results of the direction of motion illusion in the first part of the experiment were inconsistent between the participants, and the inner and outer tendon vibration did not always have the opposite effect. Multiple reasons can be considered for the results as reviewed in Section \ref{LR:TV} such as some activity in neighboring or antagonistic muscles, individual differences in anatomy, and cognitive factors. Controlling all of these factors may not be practical with a simple set up such as using a band-type device. Moreover, because we used the band-type device like some previous studies (e.g., \cite{le2020influence, ushiyama2022increasing}, the vibration induced by one side might more likely affect the other side. Additionally, we considered that the experiment setup having no restriction on user movement unlike in most existing studies would increase the difficulty of controlling the effect. Because of the freedom of the movement, some participants might experience motion illusion in the direction of lengthening the muscle, whereas others might experience TVR inducing the motion in the direction of straining the muscle. Furthermore, the tendon vibration illusion can be said to be strongly biased because of top-down effects such as knowledge and expectations \cite{taylor2017muscle, thyrion2009perceptual}; therefore, the motion illusion may more likely vary in a practical setup where users can freely move their body or while they are exerting larger forces than those required to hold the lightweight controllers used in this study. Nonetheless, a correlation was found between the amount of motion illusion caused by inner and outer tendon vibration, suggesting that at least consistent individual differences exist in the illusion of motion to tendon vibration. Taking the discussions into account, it was suggested that the use of tendon vibration as noise on somatosensory information might be more practical for motion illusion than precisely controlling the effect, with the current simple setup.

\subsubsection{How does tendon vibration affect the DT?} 
The results suggest that tendon vibration has an effect on the DT of the discrepancy between actual and visual motion, particularly with a high visual gain for which visual motion was larger than an actual motion. Participants had a higher DT with tendon vibrations for the ENLARGE group of visual gain. This indicates that tendon vibration could extend the range of applicable visual gain without users noticing the visual/physical discrepancy. Regarding the effect of tendon vibration on the DT, we mentioned two different possibilities. Of these possibilities, the second possibility explained the results the best: tendon vibration functions as noise on somatosensory sensation, and participants noticed the discrepancy between visual and somatosensory information less. If the first possibility was true, the DT should decrease for inner tendon vibration compared with the control condition in the ENLARGE gain group. However, the results indicated that the DT increased in both inner and outer tendon vibration. Moreover, the results showed that no positive correlation exists between the direction of motion illusion induced by tendon vibration during rest and whether DT increases compared with the control condition. Therefore, the second possibility seems to fit these results the most.

The fact that the effect of "all" tendon vibration on DT seems not so large suggests either that the effects of inner and outer tendon vibrations cancel each other \cite{gilhodes1986perceptual, gonzales2014short}, or that the amount of the effect of tendon vibration is affected more by the amplitude of each vibration than the total amplitude of the vibrations. Although it is unclear which of the two explanations is correct based on the present experiment alone, the latter explanation is more likely to be correct, given that there was no correlation between whether the vibration is inner or outer and the direction of motion illusion.

The effect of tendon vibration was larger with the ENLARGE condition than REDUCE condition, and our results were consistent with previous results reporting that people were more sensitive to the visual/kinesthetic discrepancy in the REDUCE visual gain than ENLARGE visual gain \cite{burns2006perceptual}. 
%For our results, it might be explained by the effect of muscle construction due to the pseudo-haptics. When vibration is applied during voluntary contraction, only the portion of the spindle endings not yet activated by the fusimotor drive that is activated by the vibration is perceived as a signal of muscle stretch. Thus, as the intensity of muscle contraction increases, the rate of the motion illusion produced by vibration decreases proportionally \cite{mccloskey1973differences}. It is also known that people unconsciously try to grip with an appropriate grip force for a weight \cite{flanagan1997role}, and that pseudo-haptic perception can also affect grip force \cite{sarlegna2010delayed}. From this, it is possible that the increase in perceived load in REDUCE caused muscle contraction, which in turn weakened the effect of tendon vibration. 
One possible reason of this result might be that, in our experiment, the amount of visual motion was less with REDUCE visual gain (60 degrees) than with ENLARGE visual gain (120 degrees). Therefore, although the physical motion range was the same in the REDUCE and ENLARGE conditions (60 to 120 degrees), the visual information in the REDUCE condition may affect the motion perception less. Consequently, although the motion perception relied more on the visual information owing to the noise effect of the tendon vibration, its effect on the DT was small in REDUCE visual gain.

%%%%%%%%%%%%%%%%%%%%%%%%%%%%%%%%%%%%%%%%%%%%%%%%%%%%%%%%%%%%%%%%%%%%%%%%%%%%%%%%%%%%%%
%XP2
%%%%%%%%%%%%%%%%%%%%%%%%%%%%%%%%%%%%%%%%%%%%%%%%%%%%%%%%%%%%%%%%%%%%%%%%%%%%%%%%%%%%%%

\section{XP2: Influence of tendon vibration on resolution of pseudo-weight perception}

In this experiment, the JND of pseudo-weight perception induced by motion gain with and without inner tendon vibration was investigated. The first experiment indicated that the tendon vibration could extend the range of applicable visual gain. However, considering the possibility that tendon vibration functions as noise on somatosensory information, tendon vibration may degrade the sensitivity of pseudo-weight perception. Therefore, the research question of this experiment was whether the tendon vibration affected the sensitivity of pseudo-haptic weight perception. Note that if the tendon vibration improves or does not change the JND compared with the no-vibration condition, we can conclude that tendon vibration can effectively increase the range of pseudo-haptic weight perception.

\subsection{Methods}
The equipment and virtual environment was the same as the first experiment.

\subsubsection{Task}
The experiment was conducted using the constant stimuli method. The task involved comparing the weights of two different virtual dumbbells: the reference and comparison dumbbells. First, participants lifted the first virtual dumbbell. The lifting task was the same as that in the first experiment for the ENLARGE gain group (including 6 seconds waiting phase). When the virtual dumbbell reached the blue bar, the dumbbell disappeared and the other dumbbell appeared at the initial position. Then, participants lifted the second dumbbell to the blue bar. Subsequently, a question panel was presented in front of them, and they had to answer the question "Which dumbbell did you feel heavier?" with "the former" or "the latter." They used the left VR controller to answer. Subsection 4.1.2. and 4.1.4 provides the detailed condition and flow, respectively.

\subsubsection{Conditions}
There were two conditions regarding the vibrations: with and without inner tendon vibration. The vibration conditions of the reference and comparison were the same. This means that if the reference is without vibration, the comparison is without vibration, and if the reference is with vibration, the comparison is with vibration. In addition, the reference dumbbell had a motion gain of 1.5, and the comparison dumbbell had motion gains of -30\%, -15\%, -8\%, +8\%, +15\%, or +30\% of the reference gain (that is, motion gains of 1.05, 1.28, 1.38, 1.62, 1.73, 1.95, respectively). These values were determined through the pilot test. Note that several motion gains exceeded the DT. Our pilot test confirmed that participants could experience a pseudo-weight perception for these gains while they noticed the motion modification. We considered only the inner tendon vibration because it was expected to have the biggest effect on the results, considering the results of the first experiment.

\subsubsection{Collected data}
Each participant answer was measured to calculate the JND of pseudo-weight perception. Furthermore, questionnaires regarding tiredness, confidence of the answer, strategy to answer, and impression of vibration were asked.

\subsubsection{Procedure}
Ten participants (7 males and 3 females in their 20s) participated. We would like to note that it is known that about 10 participants are enough to evaluate pseudo-haptic perception by the method of constant stimuli (e.g., \cite{stellmacher2022triggermuscle, weiss2023using}). Moreover, some of them were the same participants as the ones in the first experiment. The flow was almost the same with the first experiment until the first part of the experiment including the consent form. The difference was that the calibration and measurement were conducted only for the inner tendon vibration. Subsequently, participants continued to the main part of the constant stimuli method. There were 5 blocks in total. Within a block, all combinations of the two vibration conditions and six motion gains of the comparison condition were presented considering the presenting order of the reference and comparison dumbbells. Therefore, each block included 24 trials, and each comparison condition was compared to the reference condition ten times. At the end of the experiment, participants removed all of their equipment and filled in a questionnaire. Subsequently, they were remunerated with an Amazon gift card of 20 euro for their participation.

\subsubsection{Research Question}
The research question of this experiment is as follows:

\begin{itemize}[\IEEEsetlabelwidth{\textrm{[Q2-1]}}]
    \item [\textrm{[Q2-1]}] Does tendon vibration degrade the sensitivity of pseudo-haptic weight perception?
\end{itemize}

\begin{figure}[t]
 \centering
 \includegraphics[width=\columnwidth]{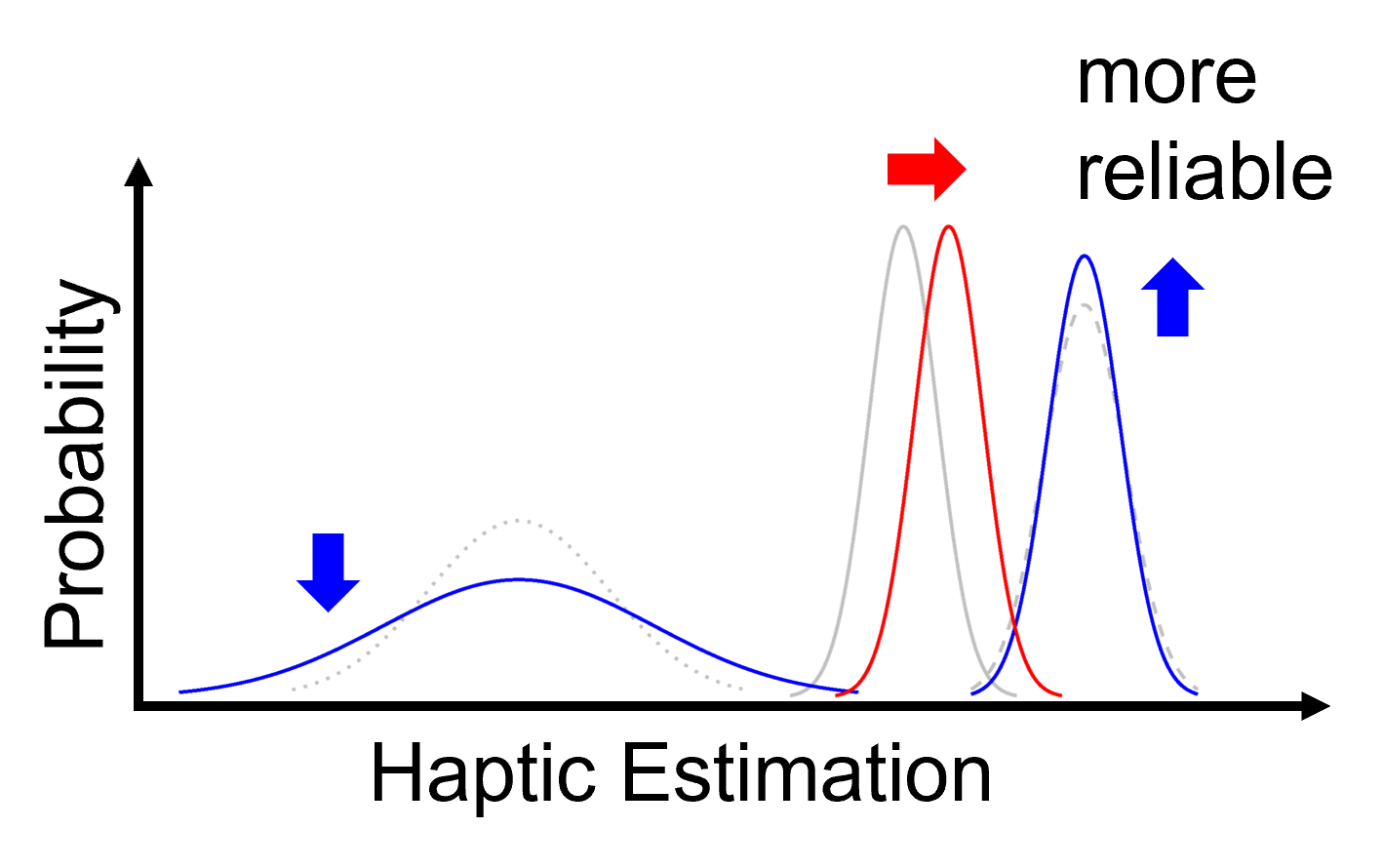}
 \caption{The figure shows the concept figure of MLE with noise on haptic information and increment of the reliability of visual information. The increment of the reliability of visual information decreases its variances and makes the final perception have fewer variances and shift to the value estimated by the visual information.}
 \label{MLE_noise_attention}
\end{figure}

As depicted in Fig.\ref{MLE_noise}, the variance of the somatosensory sensation increases if the tendon vibration functions as noise, increasing the variance of the final perception: pseudo-weight perception. In this case, the JND increases. However, we hypothesized that participants would rely more on visual cues because of the noise on somatosensory information and would then decrease the variance of the visual sensation (Fig. \ref{MLE_noise_attention}). If this occurs, the JND would not be necessarily increased.

\subsection{Results: JND of pseudo-weight perception}
\begin{figure}[t]
 \centering
 \includegraphics[width=\columnwidth]{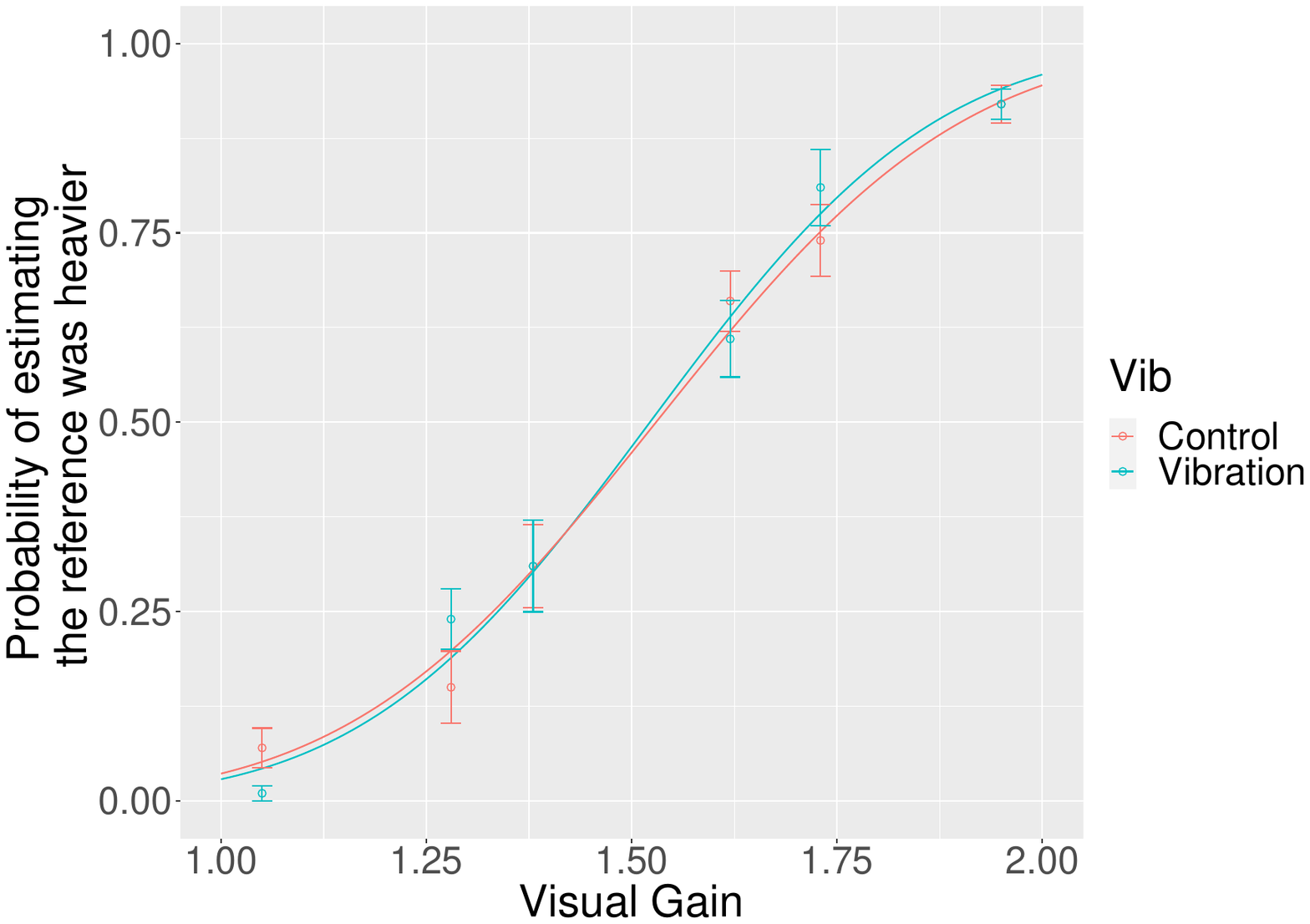}
 \caption{The figure shows the psychometric curves of the averaged results. It plots the average percentage of answers in which participants considered the reference dumbbell heavier than the comparison dumbbell. The error bar indicates the standard error. Note that Control indicates the condition without vibration.}
 \label{PsychoCurve}
\end{figure}

All participants successfully confirmed to experience motion illusion in the first part of the experiment. The Probit analysis \cite{finney1971probit} was conducted for the results of each participant, which calculated the parameters of the best-fitting cumulative normal function. Then, we computed the JND for each vibration condition as a half of the distance between the points of 25\% and 75\% on the psychometric curve for each participant. These values were 0.20 $\pm$ 0.02 for the condition without vibration (control) and 0.19 $\pm$ 0.01 for the condition with the inner tendon vibration (vibration). The JND of the vibration condition was approximately 5\% smaller than that of the control condition, but a t-test did not show the significant difference. Figure \ref{PsychoCurve} shows the psychometric curves calculated with the averaged values of all participants' results.

\subsection{Discussion}
The statistical analysis of the JND results suggested that it cannot be said that the tendon vibration degrade the JND. Here, it is complicated to statistically confirm that tendon vibration does not degrade the JND, but still, we can assume that the tendon vibration does not have a large effect on the JND. Tendon vibration is known to degrade proprioceptive responsiveness \cite{bock2007method, brun2015anchoring} and this should induce an increment of the variance of somatosensory information and thus induce that of final perception, that is, pseudo-weight perception. However, considering the results obtained, the variance of visual information may decrease. Regarding this, as mentioned above, we hypothesize that this is because participants relied more on visual information, and this decreases the variance of the information \cite{johnston1986selective}.

However, there is other possible explanations for these results. First one is that tendon vibration may have small effect as noise on haptic weight perception. Moreover, it is also possible that the effect of the decrement in the reliability of the haptic motion cues becomes small in the sensory processing for weight perception. Here, one thing to note is that the noise effect investigated in the first experiment was on haptic motion cue but the JND investigated in the second experiment was on haptic weight perception. The path of the effect of motion cue on force/weight estimates is yet unclear and can be complicated. Therefore, the effect of the noise may become small during the long way of sensory processing with other cues. Although clarifying the mechanism of pseudo-haptics is not the main objective of this study, future work can investigate this point.

%%%%%%%%%%%%%%%%%%%%%%%%%%%%%%%%%%%%%%%%%%%%%%%%%%%%%%%%%%%%%%%%%%%%%%%%%%%%%%%%%%%%%
%XP3
%%%%%%%%%%%%%%%%%%%%%%%%%%%%%%%%%%%%%%%%%%%%%%%%%%%%%%%%%%%%%%%%%%%%%%%%%%%%%%%%%%%%%

\section{XP3: Influence of tendon vibration on weight perception measured by pseudo-weight induced by visual motion gain}
This experiment investigated the PSE between the pseudo-weight perception induced by the inner tendon vibration and that by motion gain. The first and second experiments clarified the effectiveness of using tendon vibration as noise on haptic sensation for enhancing pseudo-haptics. However, through the first and second experiments and our informal test, the side effect of tendon vibration was observed, that is, it seems to function as an additional haptic weight cue, increasing a sense of weight while lifting a virtual object in VR.
As mentioned in Section \ref{LR:TV}, the effect of tendon vibration on force and weight perception has been investigated in previous studies. However, first, as the results regarding the motion illusion in the first experiment were different from those obtained in previous studies, we considered that investigating the effect on weight was also important using our practical setup. In addition, to the best of our knowledge, this study is the first to measure the effect of tendon vibration on the sense of weight by comparing the pseudo-weight perception induced by motion gain. The objective of this experiment was to investigate the PSE between the sense of weight induced by tendon vibration and by motion gain. The knowledge of the PSE leads the design guide of the use of tendon vibration with pseudo-haptics technique concerning the sense of weight.

\subsection{Methods}
The equipment and virtual environment was the same as the first and second experiments.

\subsubsection{Task}
The experiment was conducted following a staircase design. The objective was to determine the PSE between the sense of weight induced by the inner tendon vibration and that by a motion gain. The inner tendon vibration was used considering the results of the first experiment that indicated that the inner tendon vibration was most effective to the motion perception for our lifting task. As a task of the staircase method, participants lifted a reference and comparison dumbbell independently, and then answered the two-alternative forced choice (2AFC) question “Which is heavier?” with "the former," or "the latter." The lifting task was the same as that of the first and second experiments (including 6 seconds waiting phase).

\subsubsection{Conditions}
There were primarily 2 groups of staircase methods: vib-ref and con-ref. In the vib-ref group, the reference condition was a motion gain of 1.0 with inner tendon vibration. Here, the comparison condition was without vibration and with a motion gain that was one of ten steps from 0.4 to 1.0. In the con-ref group, the reference condition was without vibration and with a motion gain of 1.0. Then, the comparison condition was with inner tendon vibration and a motion gain that was one of ten steps from 1.0 to 2.5. Our pilot test determined these values. In the vib-ref group, the PSE indicates a motion gain that induces the same weight perception as the inner tendon vibration. In the con-ref group, the PSE indicates the motion gain required to cancel the sense of weight induced by the inner tendon vibration.

\subsubsection{Collected data}
Each motion gain at a turning point was measured to calculate the PSEs by averaging them.

\subsubsection{Procedure}
Twenty participants (12 males and 8 females in their 20s) participated in the experiment. We would like to note that some of them were the same participants as the ones in the first or second experiment. The procedure before the main task was the same as that in the first and second experiments, including the consent form, calibration of the vibrator positions and measurement of motion illusion. The main part was conducted with the staircase method, and the vib-ref and con-ref groups were the two blocks. Both groups had two series where one began with a minimum motion gain of 0.4 for vib-ref and 1.0 for con-ref, and the other with a maximum gain of 1.0 for vib-ref and 2.5 for con-ref. The gain of the comparison dumbbell was increased or decreased in an interval of 0.06 for vib-ref group or 0.15 for con-ref group following the participants' answers. When participants answered that the comparison dumbbell was heavier, the motion gain increased by one step. When participants answered that the reference dumbbell was heavier, the motion gain decreased by one step. A block of stair case methods ended when both series reached the 5th change of the direction on increasing or decreasing the gain. In addition, the motion gain did not cross over the minimum or maximum gain. If the motion gain was to cross over the limit, it remained at the same gain, and the count of turning points increased. The presenting order of reference and comparison conditions were randomized. In addition, the order of blocks was counterbalanced. The participants were remunerated with an Amazon gift card of 5 euro for their participation.

\subsection{Results: PSE of pseudo-weight perception between tendon vibration and motion gain}
\begin{figure}[t]
 \centering
 \includegraphics[width=\columnwidth]{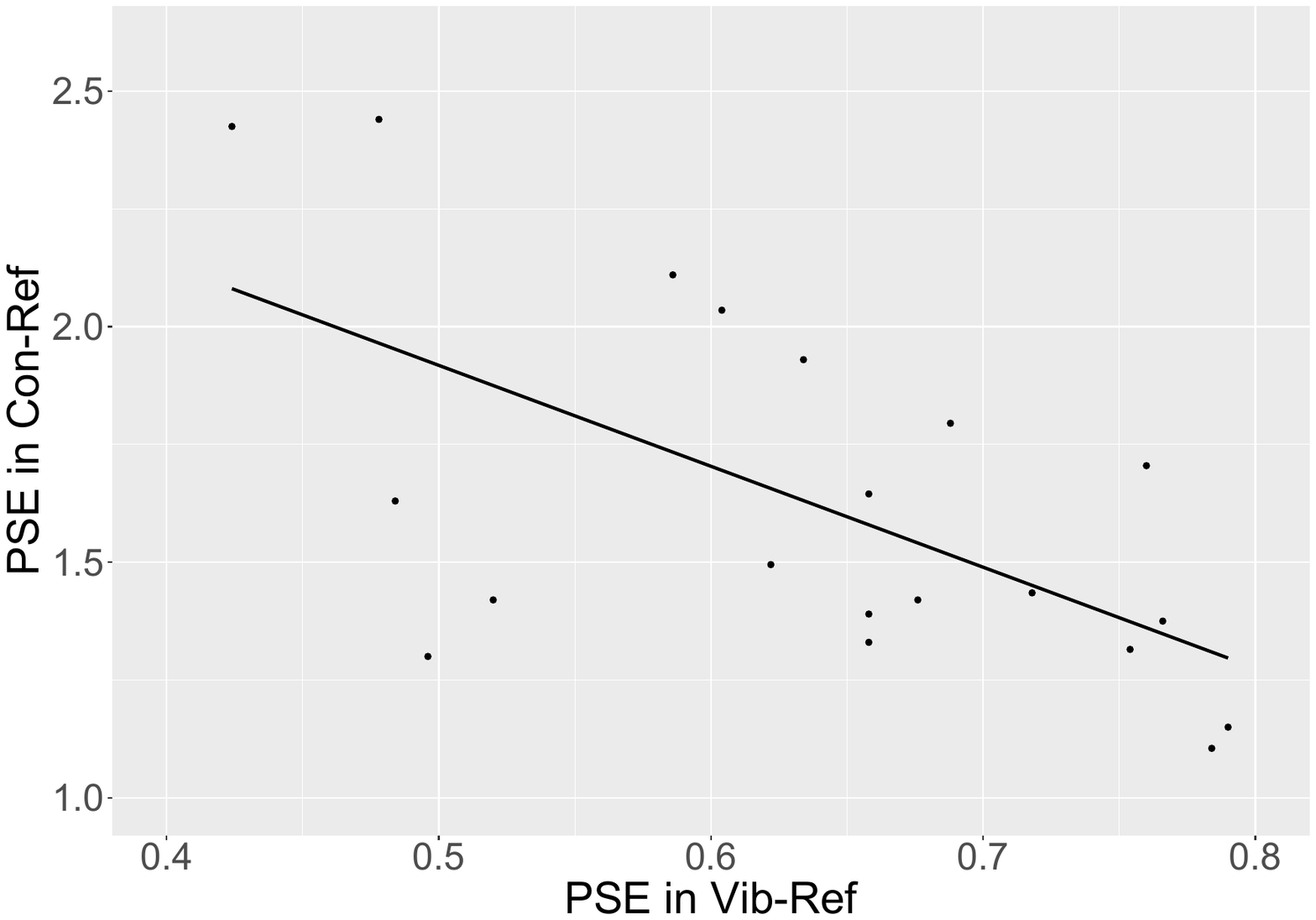}
 \caption{The figure shows the plots of the PSE results of each participant in the vib-ref and con-ref groups.}
 \label{Plot_Correlation}
\end{figure}

All participants successfully confirmed experiencing motion illusion in the first part of the experiment. The 10 motion gains at the turning points in a block were averaged for each participant to compute the PSE. Table \ref{PSE} indicates the PSE results. The PSE was $0.64 \pm 0.22$ in the vib-ref group and $1.62 \pm 0.76$ in the con-ref group (each PSE was $average \pm 2\sigma$). Here, the results of the PSE had large variances beyond the JNDs obtained in XP2. Then, we analyzed the correlation between the two PSE results to determine if they were consistent within each participant. Figure \ref{Plot_Correlation} shows the plots of PSE results of each participant in the vib-ref and con-ref groups. The Pearson correlation test was conducted for the relationship between the results of the PSE in the vib-ref and con-ref groups. Consequently, we determined a significant negative correlation ($r=-0.61, p=0.0042$). In addition, the same correlation test was conducted for the relationship between the absolute value of the intensity of motion illusion at rest and each PSE result. Then, a significant positive correlation was found between motion illusion and the PSE of the con-ref group ($r=0.68, p=0.00093$).

%\begin{figure}[t]
% \centering
% \includegraphics[width=\columnwidth]{Figures/PSE_Combined.png}
% \caption{PSE of pseudo-weight perception between the inner tendon vibration and motion gain. Error bar indicates 2SD.}
% \label{PSE}
%\end{figure}

\begin{table}[t]
\caption{The figure shows the PSE of pseudo-weight perception between inner tendon vibration and motion gain.}
\label{PSE}
\begin{center}%\small
\begin{tabular}{|c||c|c|}\hline 
Condition & PSE ($average\pm2\sigma$) \\ \hline \hline
vib-ref   & $0.64 \pm 0.22$       \\ \hline
con-ref   & $1.62  \pm 0.76$       \\ \hline
\end{tabular}
\end{center}
\vspace*{-3mm}
\end{table}

\subsection{Discussion}
The results suggest that the inner tendon vibration increases the sense of weight and that the amount of the sense of weight increased by tendon vibration was the same as the amount increased by a visual gain of 0.64 and the one decreased by a visual gain of 1.62. Studies mentioned in the related work section have noted that tendon vibration increases the signal from the muscle spindle and/or tendon organ, and this might lead to an increment in the sense of weight. Here, the PSE results from the two groups showed a symmetrical relationship, indicating that the tendon vibration's effect of increasing the sense of weight would remain almost constant across different visual gains. This should be clarified in future work where the weight perception during synergistic use of tendon vibration and pseudo-haptics will be evaluated. However, the results had large variances. As shown in Fig. \ref{Plot_Correlation} and the results of the correlation test, participants had consistent PSE results between the vib-ref and con-ref groups. This indicates that the effect of the vibration on pseudo-weight perception has individual differences. The results of the correlation test between the intensity of motion illusion at rest and PSE of the con-ref group support this.

Moreover, these PSE results indicate that the effect of tendon vibration as noise on haptic weight information is not strong. If the tendon vibration functions as noise on haptic force information, the effect of visual motion gain should become stronger and users should feel lighter at a gain above 1. However, the results did not show this, suggesting that the effect of tendon vibration as an additional haptic force cue is stronger than that of noise for haptic force information. Here, one thing to note is that what the results of the first experiment indicate is that the tendon vibration functions as noise on a haptic "motion" cue.

\section{General Discussion}
This study investigated whether tendon vibration can extend the effect of the pseudo-haptic technique. Our results suggest that the inner tendon vibration while lifting a virtual object can extend an applicable motion gain without users detecting visual/physical discrepancy (XP1) with the similar resolution of pseudo-weight perception as without vibration (XP2). Our first experiment suggested that an unnoticeable, strict applicable gain was from 0.71 to 1.41 without tendon vibration and 0.74 to 1.53 with inner tendon vibration. This indicates that the tendon vibration extends the range of unnoticeable visual gain by about 13\%. In addition, the second experiment showed that the JND of pseudo-weight perception was 0.20 and 0.19 for each condition. Here, we could roughly conclude that visual motion gain could present at least four levels of weight perception (by $(1.41-0.71)/0.20$), and visual motion gain with inner tendon vibration could present at least five levels of weight perception (by $(1.53-0.74)/0.19$). This effect indicates that tendon vibration could increase the number of presentable levels of pseudo-weight perception. However, we would like to note that the JND was investigated around a reference gain of 1.5 and may be different if investigated with different reference gain. This should be clarified in future work. To the best of our knowledge, this study is the first to show the possibility of using tendon vibration as noise on somatosensory information (specifically, haptic motion cue) to increase the contribution of visual information in the sensory integration processing.

\begin{figure}[t]
 \centering
 \includegraphics[width=\columnwidth]{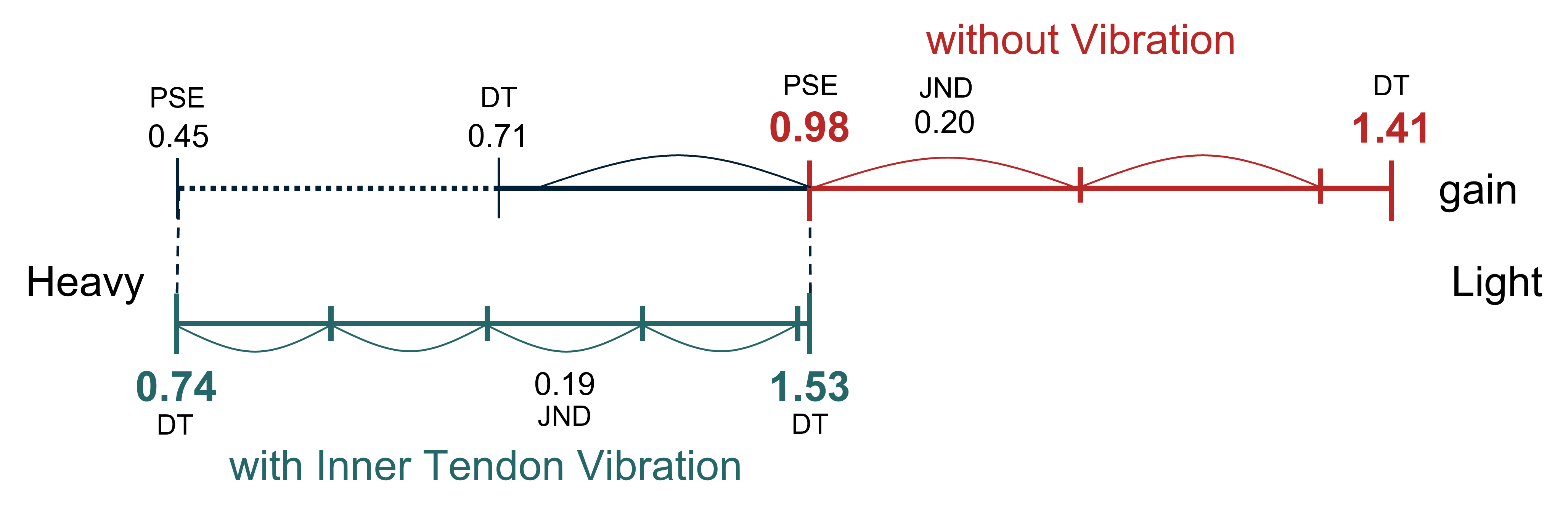}
 \caption{The figure shows the summary of all results. Here, the horizontal bars indicate the range of gain where pseudo-weight perception can be presented without users noticing the visual/physical discrepancies. The top bar indicates the range of the condition without tendon vibration and the bottom bar indicates that with tendon vibration. The bars are scaled to make them comparable in the axis of the same pseudo-weight perception.
 The pseudo-haptic technique can present four steps of different weight perceptions without the tendon vibration (the range: 0.71 - 1.41; the resolution: 0.20) while it can present five steps with the tendon vibration (the range: 0.74 - 1.53; the resolution: 0.19). The most efficient manner of combining visual motion gain and inner tendon vibration is using a gain of 0.74 to 1.53 with the tendon vibration and 0.98 to 1.41 without vibration. Then, the pseudo-haptic technique can present 7 steps in total. }
 \label{SummaryOfResults}
\end{figure}

However, it was clarified that the tendon vibration has a side effect. Although it functions as noise on a haptic motion cue, it also functions as an additional haptic weight cue. To investigate this effect, this study is the first to compute the PSE of pseudo-weight perceptions between tendon vibration and motion gain (XP3). With the PSE obtained in our experiment, developers can design a wider range of pseudo-weight perception by combining tendon vibration and motion gain. Fig.\ref{SummaryOfResults} summarizes our results. Practically, as the PSE was 0.64 and using the results of the first experiment, we could compute that the presentable range of weight perception with tendon vibration and gain is to be theoretically the same as 0.45 ($=0.64*0.74$) to 0.98 ($=0.64*1.53$) without tendon vibration. Therefore, considering the resolution (results of XP2), the most efficient manner of combining techniques is to use 0.74 to 1.53 of gain with tendon vibration and 0.98 to 1.41 without tendon vibration. Then, the technique can present at least seven levels of weight perception (by $(1.53-0.74)/0.19+(1.41-0.98)/0.20$). The technique should nearly double the capacity for simulating virtual weights: four steps without vibration, and up to seven steps exploiting vibrations.

One of the limitations of this study is that the effect of tendon vibration has individual differences. Therefore, future work can investigate this individual difference in more detail and the calibration system of the map between the effects of tendon vibration and motion gain for each user. Furthermore, we would like to note that our experiments do not negate the potential impact of vibratory feedback on other factors, such as skin, muscles, and other neighboring muscles/tendons, like other previous work (e.g. \cite{taylor2017muscle, le2020influence, ushiyama2021modulation}). Therefore, how to fit the vibrators to more accurately stimulate the exact tendons or how to avoid other unintended effects should be discussed in future work. For example, future work may investigate the effects with other vibration frequencies such as 200 Hz or higher where skin stimulation is performed but tendon vibration is not. Moreover, as one of the general limitations in the application of tendon vibration, we need to note that the motion illusion is said to occur several seconds after the start of the stimuli. However, the results obtained in the study suggest the use of tendon vibration not as precise control of motion illusion but as noise or additional force/weight cue on somatosensory information. To our knowledge, it is not confirmed if the pre-vibration is required for those effects and thus future study can investigate this point. Moreover, the future study below may shorten the required pre-vibration time.

Future research may explore additional vibration parameters and interaction designs that can further enhance the effect. One possible method for enhancing the effect is to utilize the aftereffect of tendon vibration \cite{kito2006sensory}. For example, to increase the effect of motion illusion, activate one side of the tendon vibration until before lifting a virtual object and then switch it to the other side of the tendon vibration immediately after lifting. Moreover, combination of different frequencies of vibrations for agonist and antagonist tendons would be one of the interesting topics for future work as suggested by Gilhodes et al., \cite{gilhodes1986perceptual}. In addition, in the second and third experiments, several visual gains were used above the DT. The DT of the visual/physical motion discrepancy obtained in the first experiment is the strictest criterion for an applicable visual gain. However, we believe that manipulating the weight perception with gains above the DT is possible and that the practical range of the applicable visual gain is greater than the DT obtained in the first experiment. First, in our experiment and in other psychological experiments that investigate this type of DT, the threshold would be stricter than in practical use case because the tasks require participants to focus on the DT, thereby improving sensitivity to it. Moreover, we consider that a visual gain with users feeling uncomfortable or without sensory integration occurring because of a large visual/physical motion discrepancy is not the same as the DT. Some studies have positively applied visual gain above the DT and succeeded in presenting a greater pseudo-weight perception \cite{rietzler2018breaking}. However, a method for measuring the range of this practical visual gain has not yet been established. Therefore, another study could propose an experimental method for investigating this practical range of pseudo-haptics with tendon vibration. Finally, the limitation of our use of tendon vibration is that it emits noise and may impair the VR experience. Although no participant mentioned that they felt uncomfortable or annoyed by the sound or vibration, using noise-canceling headphones and avoiding long-duration uses may create better experiences.

Moreover, considering that the experimental results suggested that tendon vibration has a side effect and functions as an additional haptic force cue, future work can also investigate if there is a method that functions purely as noise to haptic cues or has smaller other effects. When noise is added to deep sensation, nerves in muscles and tendons fire. Here, it is theoretically possible to counteract the direction of the effect by approaching agonist and antagonist muscles/tendons at the same time. However, our experimental results suggested that the directional perception of motion or force induced by tendon vibration does not relate to the effect of the additional haptic weight cue. Therefore, it may be considered that the absolute input value of the muscles/tendons, regardless of their position, functions as the additional haptic weight cue. If it is true, it is difficult not to have the effect of additional haptic weight (force) cues when adding noise on deep sensation even though a different approach such as electrical muscle stimuli (EMS) is used instead of tendon vibration.
One possible way to avoid such an unintended effect is to desensitize the target muscles/tendons by constantly presenting vibration throughout the experience of applications \cite{hultborn1996mechanism, wood1996influence}. 
It is suggested that presenting tendon vibration longer than a certain period of time, the vibrated muscles/tendons start to be desensitized and the muscle/tendon exertion is underestimated \cite{luu2011fusimotor}. Then, if the nerves of muscle/tendon can be saturated in this way, the stimuli on muscles/tendons by using tendon vibration or EMS may not result in an additional haptic weight (force) cue. However, there are several things to be noted and investigated in future work. First, as discussed in Section \ref{LR:TV}, the effect of tendon vibration is not robust and to our knowledge, there is no clear consensus on the effects of desensitization on force perception while moving. Therefore, the effect should be investigated in several different scenarios such as using different motion trajectories and moving speeds, considering individual differences. Moreover, the effect of constant vibration/EMS presentation on user experiences also needs to be investigated. 

\section{Conclusion}
This study investigated the possibility of using tendon vibration to enhance pseudo-haptic perceptions in VR during free arm motion. In particular, we evaluated the effect of tendon vibration on the DT of visual/physical motion discrepancy (XP1), the JND of pseudo-weight perception induced by visual gain (XP2), and the sense of weight by computing the PSE with the pseudo-weight induced by visual gain (XP3). The results of the first and second experiments show the possibility of a novel approach that leverages tendon vibration as noise on somatosensory information  (specifically, haptic "motion" cue) to enable the sensory integration to rely more on a visual cue. Specifically, it was suggested that tendon vibration can increase the range where users can obtain pseudo-weight perception without noticing the visual/physical motion discrepancy by 13\%. 
In addition, the third experiment investigated the side effect of tendon vibration by measuring the PSE. The results suggest that the effect of tendon vibration as noise for a haptic force/weight cue would be weaker than the effect as an additional haptic force/weight cue. This means that the tendon vibration increases weight perception. Future study would investigate the way of alleviating this side effect. Still, the PSE obtained in the third experiment helps developers design a wider range of pseudo-weight perception.

\section*{Acknowledgments}
This work was partially supported by the MEXT Grant-in-Aid for Scientific Research (S) (19H05661), JSPS Grant-in-Aid for Scientific Research (A) (21H04883), and Grant-in-Aid for JSPS Fellows (21J12284).

\ifCLASSOPTIONcaptionsoff
  \newpage
\fi



\bibliographystyle{IEEEtran}
\bibliography{main}

%\begin{thebibliography}{1}
%\end{thebibliography}


\vspace*{-3\baselineskip}

\begin{IEEEbiography}[{\includegraphics[width=1in,height=1.25in,clip,keepaspectratio]{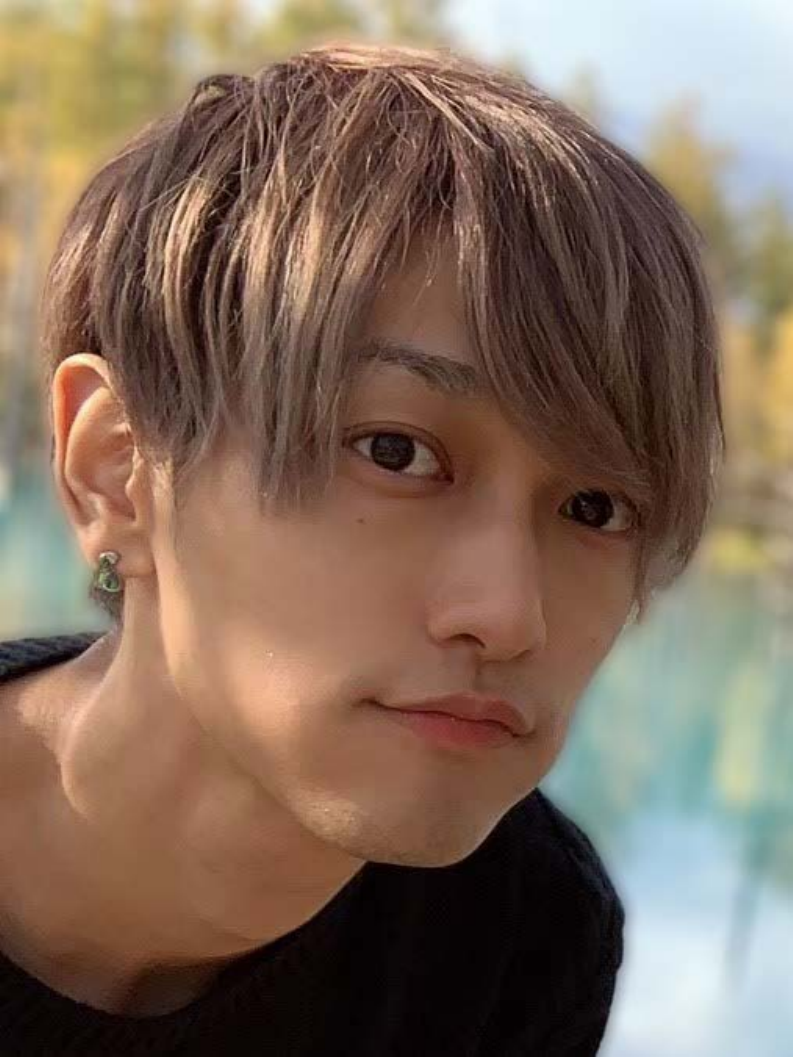}}]{Yutaro Hirao}
received his B.S. and M.S. in engineering from Waseda University (2018 and 2020) in Japan. He is currently working toward
the PhD degree in engineering from the University of Tokyo. His research topics are mainly virtual reality (VR) and cross-modal interaction.
\end{IEEEbiography}

\vspace*{-3.4\baselineskip}

\begin{IEEEbiography}[{\includegraphics[width=1in,height=1.25in,clip,keepaspectratio]{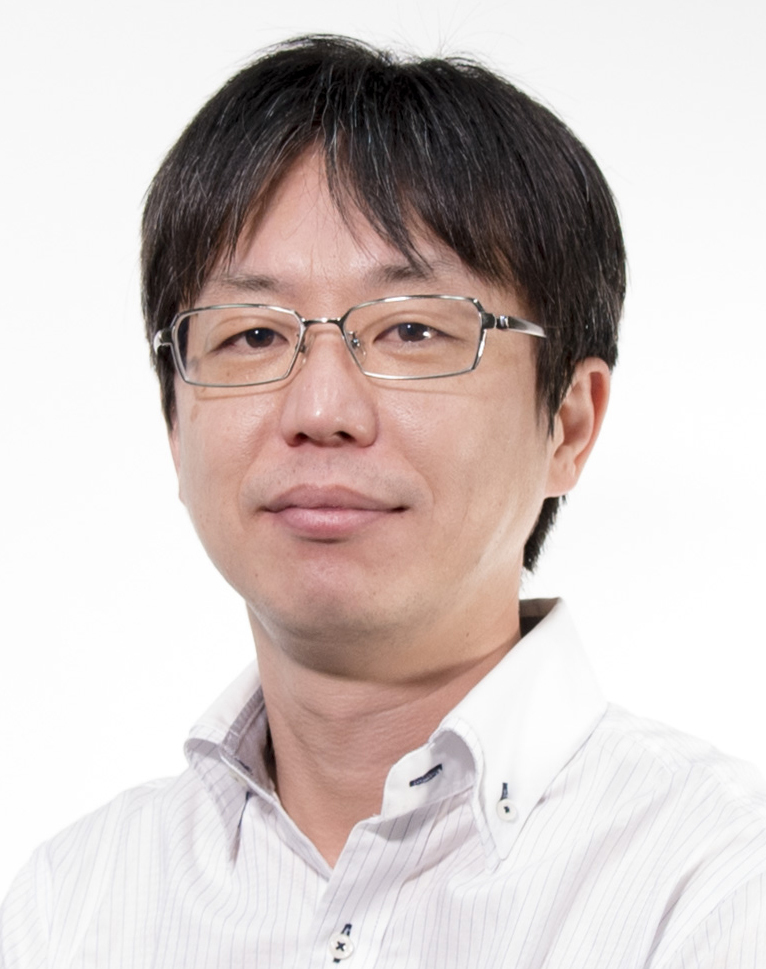}}]{Tomohiro Amemiya}
 is an associate professor at the Graduate School of Information Science and Technology, the University of Tokyo. He received BS and MS degrees in mechano-informatics from the University of Tokyo in 2002 and 2004. From 2004 to 2019, he was a research scientist at NTT Communication Science Laboratories. He received his Ph.D. from Osaka University in 2008. His research interests include somatosensory perception and human-computer interaction technologies.
\end{IEEEbiography}

\vspace*{-3.4\baselineskip}

\begin{IEEEbiography}[{\includegraphics[width=1in,height=1.25in,clip,keepaspectratio]{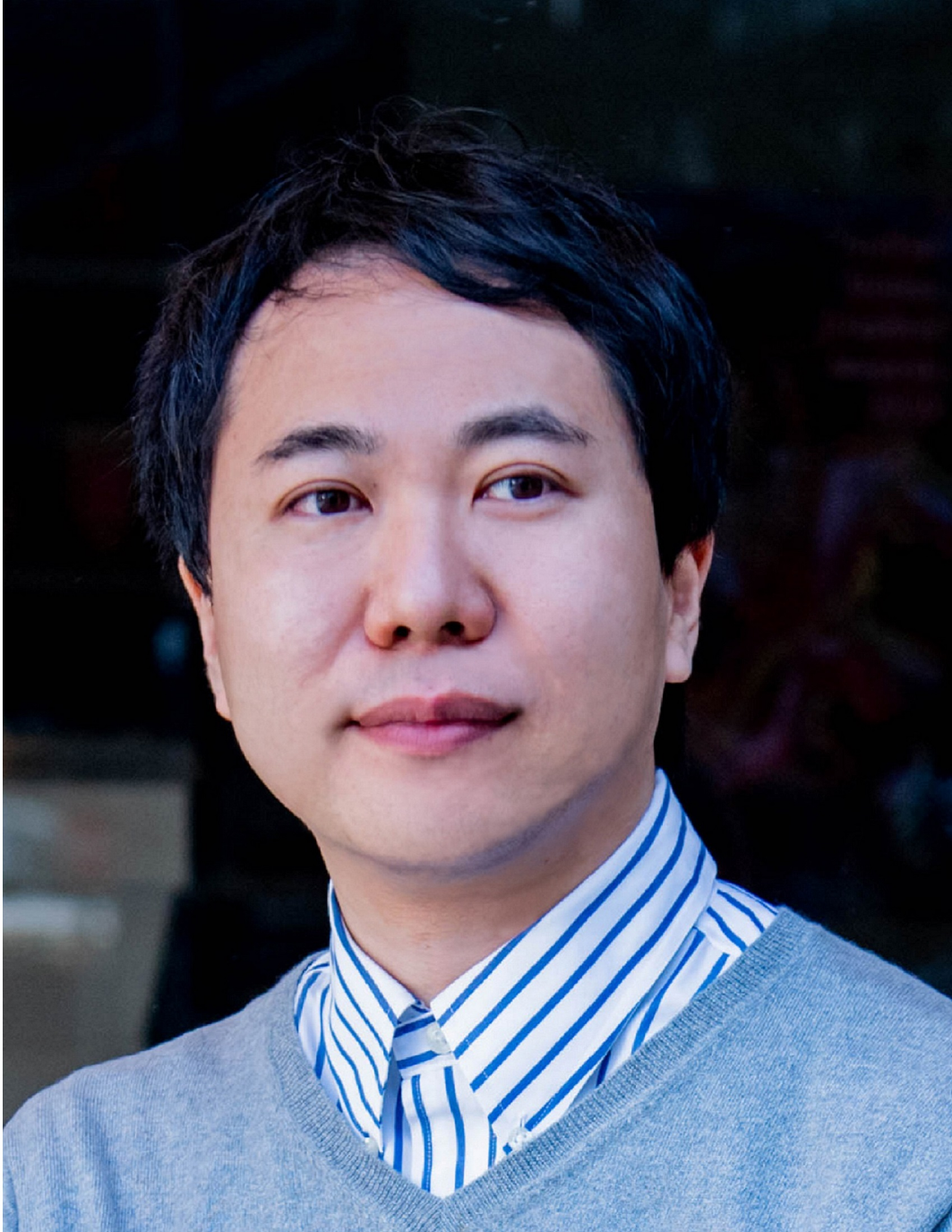}}]{Takuji Narumi}
is an associate professor at the Graduate School of Information Science and Technology, the University of Tokyo. His research interests broadly include perceptual modification and human augmentation with virtual reality and augmented reality technologies. He received BE and ME degree from the University of Tokyo in 2006 and 2008 respectively. He also received his Ph.D. in Engineering from the University of Tokyo in 2011.
\end{IEEEbiography}

\vspace*{-4\baselineskip}

\begin{IEEEbiography}[{\includegraphics[width=1in,height=1.25in,clip,keepaspectratio]{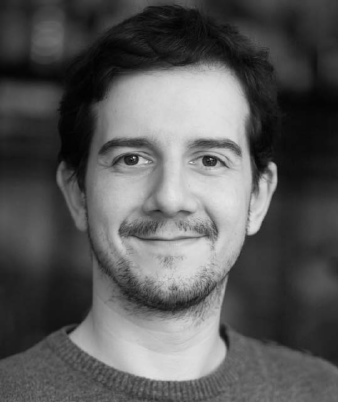}}]{Ferran Argelaguet}
is an Inria research scientist at the Hybrid team (Rennes, France) since 2016. He received his PhD degree from the Universitat Polit\`ecnica de Catalunya (UPC), in Barcelona, Spain in 2011. His main research interests include 3D user interfaces, virtual reality and human-computer interaction. He was program co-chair of the IEEE Virtual Reality and 3D User Interfaces conference track in 2019 and 2020, and the journal track in 2022.
\end{IEEEbiography}

\vspace*{-4\baselineskip}

\begin{IEEEbiography}[{\includegraphics[width=1in,height=1.25in,clip,keepaspectratio]{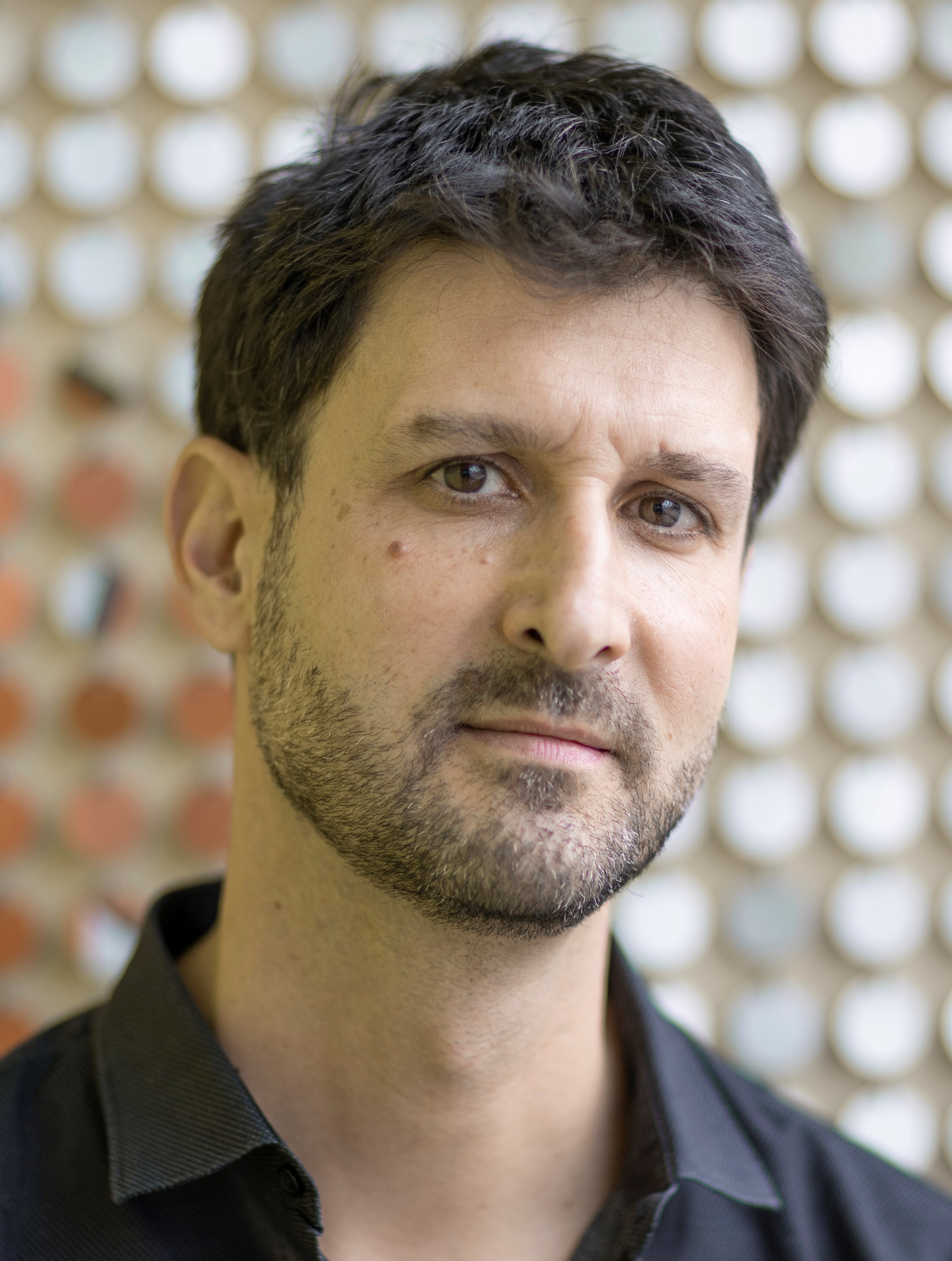}}]{Anatole Lécuyer}
is director of research and head of Hybrid team at Inria, Renne, France. He is currently Associate Editor of IEEE Transactions on Visualization and Computer Graphics, Frontiers in Virtual Reality and Presence. He was Program Chair of IEEE VR 2015-2016 and General Chair of IEEE ISMAR 2017. Anatole Lécuyer obtained the IEEE VGTC Technical Achievement Award in Virtual/Augmented Reality in 2019.
\end{IEEEbiography}

\vspace*{-3\baselineskip}

%\newpage




%\vfill


\end{document}